\documentclass[authorversion,nonacm,acmsmall]{acmart}

\setcopyright{acmlicensed}
\copyrightyear{2018}
\acmYear{2018}
\acmDOI{XXXXXXX.XXXXXXX}

\acmJournal{JACM}
\acmVolume{37}
\acmNumber{4}
\acmArticle{111}
\acmMonth{8}

\usepackage{subcaption}
\usepackage{booktabs}
\usepackage{listings}
\usepackage{siunitx}
\usepackage[inference]{semantic}
\usepackage[linesnumbered]{algorithm2e}
\usepackage{suffix}
\usepackage[export]{adjustbox}
\usepackage{tikz}
\usepackage[varqu]{zi4}
\usepackage{refcount}

\makeatletter
\let\old@lstKV@SwitchCases\lstKV@SwitchCases
\def\lstKV@SwitchCases#1#2#3{}
\makeatother
\usepackage{lstlinebgrd}
\makeatletter
\let\lstKV@SwitchCases\old@lstKV@SwitchCases

\lst@Key{numbers}{none}{%
  \def\lst@PlaceNumber{\lst@linebgrd}%
  \lstKV@SwitchCases{#1}%
                    {none:\\%
                      left:\def\lst@PlaceNumber{\llap{\normalfont
                          \lst@numberstyle{\thelstnumber}\kern\lst@numbersep}\lst@linebgrd}\\%
                      right:\def\lst@PlaceNumber{\rlap{\normalfont
                          \kern\linewidth \kern\lst@numbersep
                          \lst@numberstyle{\thelstnumber}}\lst@linebgrd}%
                    }{\PackageError{Listings}{Numbers #1 unknown}\@ehc}}
\makeatother
\usepackage{lstlinebgrd}

\newenvironment{todo-outline}
  {\begin{itemize}}
  {\end{itemize}}

\newcommand{\sectionBreakMaybe}{}

\definecolor{framegray}{rgb}{0.9,0.9,0.9}
\definecolor{listinggray}{rgb}{0.95,0.95,0.95}

\newcommand{\pgftextcircled}[1]{
    \setbox0=\hbox{#1}%
    \dimen0\wd0%
    \divide\dimen0 by 2%
    \begin{tikzpicture}[baseline=(a.base)]%
        \useasboundingbox (-\the\dimen0,0pt) rectangle (\the\dimen0,1pt);
        \node[circle,draw,outer sep=0pt,inner sep=0.1ex] (a) {#1};
    \end{tikzpicture}
}

\newcounter{claimi}
\newenvironment{claims}{
\begin{enumerate}
\setcounter{enumi}{\value{claimi}}
}{
\setcounter{claimi}{\value{enumi}}
\end{enumerate}
}

\lstdefinestyle{error}{%
  moredelim=**[is][\color{red}]{@}{@},
}

\definecolor{ckeywords}{rgb}{0.13,0.13,1}
\definecolor{ccomments}{rgb}{0,0.5,0.5}
\definecolor{cstrings}{rgb}{0,0.5,0}
\definecolor{cwarnings}{rgb}{1,0.5,0}

\lstdefinelanguage{MCore}{%
    morekeywords={Lam,con,else,end,fix,if,in,lam,lang,let,match,recursive,sem,syn,then,type,use,utest,with,letop,letimpl,all,repr},
    otherkeywords={->,_,@,?},
    keywordstyle=\color{ckeywords},
    morekeywords=[2]{mexpr,include,never},
    keywordstyle=[2]\color{cwarnings},
    morecomment=[l][\color{ccomments}]{--},
    morestring=[b]",
    stringstyle=\color{cstrings},
    sensitive=true,
    breaklines=true,
    escapeinside={(*@}{@*)},
    stepnumber=1,
    columns=fixed, 
    showstringspaces=false,
    mathescape=true,
    breaklines=true,
    breakatwhitespace=true,
    mathescape=true,
    showstringspaces=false
  }

\lstdefinelanguage{RepCaml}[]{Caml}{
    morekeywords={letop, letimpl, letrepr, repr},
    keywordstyle=\color{ckeywords},
    stringstyle=\color{cstrings},
    commentstyle=\color{ccomments}
}

\makeatletter
\newcommand*\Reactivatenumber[1]{%
  \setcounter{lstnumber}{\numexpr#1-1\relax}
  \lst@AddToHook{OnNewLine}{%
   \let\thelstnumber\origthelstnumber%
   \refstepcounter{lstnumber}
  }%
}\makeatother

\makeatletter
\newcount\bt@rangea
\newcount\bt@rangeb

\newcommand\btIfInRange[2]{%
    \global\let\bt@inrange\@secondoftwo%
    \edef\bt@rangelist{#2}%
    \foreach \range in \bt@rangelist {%
        \afterassignment\bt@getrangeb%
        \bt@rangea=0\range\relax%
        \pgfmathtruncatemacro\result{ ( #1 >= \bt@rangea) && (#1 <= \bt@rangeb) }%
        \ifnum\result=1\relax%
            \breakforeach%
            \global\let\bt@inrange\@firstoftwo%
        \fi%
    }%
    \bt@inrange%
}
\newcommand\bt@getrangeb{%
    \@ifnextchar\relax%
        {\bt@rangeb=\bt@rangea}%
        {\@getrangeb}%
}
\def\@getrangeb-#1\relax{%
    \ifx\relax#1\relax%
        \bt@rangeb=100000
    \else%
        \bt@rangeb=#1\relax%
    \fi%
}

\makeatother

\makeatletter
\newcommand{\xRightarrow}[2][]{\ext@arrow 0359\Rightarrowfill@{#1}{#2}}
\makeatother

\newcommand{\code}[1]{\textbf{\texttt{#1}}}

\newcommand{\many}[1]{\overline{#1}}

\newcommand{\mType}{\mathit{Type}}
\newcommand{\mOp}{\mathit{Op}}
\newcommand{\mImpl}{\mathit{Impl}}
\newcommand{\mOpUse}{\mathit{OpUse}}
\newcommand{\mSol}{\mathit{Sol}}
\newcommand{\mCost}{\mathit{Cost}}

\newcommand{\costBare}{\mathit{cost}}

\newcommand{\getAt}[2]{#1_{#2}}                                   
\newcommand{\cost}[1]{\costBare(#1)}                             
\newcommand{\project}[2]{#1.#2}                                  
\newcommand{\projectTwo}[3]{\project{\project{#1}{#2}}{#3}}      

\SetKwProg{Fn}{Fn}{\string:}{}
\SetKw{Fail}{Fail}
\SetKw{In}{in}
\SetKwFor{For}{for}{do}{}
\SetKwComment{Comment}{}{}
\SetKw{Assert}{assert}
\SetKwFunction{Validate}{Validate}

\newcommand{\abstraction}{abstract type\xspace}

\newcommand{\abstractions}{abstract types\xspace}

\newcommand{\repr}{representation\xspace}

\newcommand{\reprs}{representations\xspace}
\newcommand{\Reprs}{Representations\xspace}

\newcommand{\operation}{operation\xspace}

\newcommand{\operations}{operations\xspace}
\newcommand{\Operations}{Operations\xspace}
\newcommand{\impl}{implementation\xspace}

\newcommand{\impls}{implementations\xspace}
\newcommand{\Impls}{Implementations\xspace}

\newcommand{\opUse}{op-use\xspace}

\newcommand{\opUses}{op-uses\xspace}

\newcommand{\solution}{solution\xspace}

\newcommand{\solutions}{solutions\xspace}

\newcommand{\reprVar}{representation variable\xspace}
\newcommand{\ReprVar}{Representation variable\xspace}
\newcommand{\reprVars}{representation variables\xspace}

\begin{document}

\title{Repr Types: One Abstraction to Rule Them All}

\author{Viktor Palmkvist}
\email{vipa@kth.se}
\affiliation{%
  \institution{KTH Royal Institute of Technology}
  \country{Sweden}
}
\author{Anders Ågren Thuné}
\email{anders.agren-thune@it.uu.se}
\affiliation{%
  \institution{Uppsala University}
  \country{Sweden}
}
\author{Elias Castegren}
\email{elias.castegren@it.uu.se}
\affiliation{%
  \institution{Uppsala University}
  \country{Sweden}
}
\author{David Broman}
\email{dbro@kth.se}
\affiliation{%
  \institution{KTH Royal Institute of Technology}
  \country{Sweden}
}

\begin{abstract}
The choice of how to represent an abstract type can have a major
impact on the performance of a program, yet mainstream compilers
cannot perform optimizations at such a high level.
When dealing with optimizations of data type representations, an
important feature is having extensible representation-flexible
data types; the ability for a programmer to add new abstract types
and operations, as well as concrete implementations of these,
without modifying the compiler or a previously defined library.
Many research projects support high-level optimizations through
static analysis, instrumentation, or benchmarking, but they are
all restricted in at least one aspect of extensibility.

This paper presents a new approach to representation-flexible data
types without such restrictions and which still finds efficient
optimizations. Our approach centers around a single built-in type
\code{repr} and function overloading with cost annotations for
operation implementations.
We evaluate our approach (i) by defining a universal collection
type as a library, a single type for all conventional collections,
and (ii) by designing and implementing a representation-flexible
graph library. Programs using \code{repr} types are typically
faster than programs with idiomatic representation
choices---sometimes dramatically so---as long as the compiler
finds good implementations for all operations. Our compiler
performs the analysis efficiently by finding optimized solutions
quickly and by reusing previous results to avoid recomputations.
\end{abstract}




\maketitle

\section{Introduction}\label{sec:introduction}

Much of programming in general-purpose programming languages ends up
manipulating various data structures, be they general collections
(e.g., sequences, sets, or maps) or slightly more niche data
structures (e.g., graphs, matrices, or union-find). Each of these
data structures can have many different representations (e.g.,
red-black trees and hash-maps are both maps), typically with different
performance characteristics. These differences may be asymptotic in
nature; they may grow arbitrarily big as input grows. Furthermore,
it is rarely the case that one representation is strictly better than
another in all cases; it depends on which operations are used and how
often, as well as the size of each structure. This means that the choice
of representation is highly important for optimization, but also
difficult; it requires a holistic view of the program to be optimized,
and can change easily if the program is changed.

Manually editing programs to try out different data structure
implementations is tedious, and may require many changes across the
whole program. An obvious approach to addressing this issue is to
 automatically select representations. This idea is not new; it
dates back almost 50 years at
least~\cite{lowAutomaticCodingChoice1976}. Unfortunately, implementing
such a system is challenging, for a number of reasons:

\begin{description}
\item[Efficiency.] The whole point is to pick efficient
  representations and thereby yielding efficient programs, but factors
  impacting performance are hard to predict.
\item[Representation switching.] A particularly complicating factor is
  that the ideal representation may change part-way through the
  execution of a
  program~\cite{sherwoodDiscoveringExploitingProgram2003}. For
  example, a program might build a large data structure early during
  execution, and then switch to mostly querying it. Such workloads may
  be different enough to warrant the extra work of converting from one
  representation to another and still see a substantial performance
  gain.
\item[Flexibility.] Some representations might be equivalent for a
  particular set of operations, even if they are not equivalent in
  general. For example, checking a set for membership yields the same
  result whether the set is backed by a red-black tree or a hash-set,
  even though iterating would yield the elements in different
  orders. A program that never observes the iteration order thus
  allows for greater flexibility, which an automatic representation
  selector would ideally exploit.
\item[Overhead.] The overhead of picking representations must be low
  enough to make the trade-off worth it. This is very directly
  noticeable for approaches instrumenting running programs to
  determine representations, but also relevant for offline approaches;
  excessive compilation time or the need for long-running benchmarks
  may also be prohibitive.
\item[Extensibility.] A compiler developer cannot possibly predict all
  possible datastructures and operations that may benefit from
  automatic choice. At the same time, designing a system that allows
  for extensibility without sacrificing other desireable qualities is
  very challenging.
\end{description}

\noindent Previous approaches to automatic representation selection
are varied and include, e.g., static
analysis~\cite{lowAutomaticCodingChoice1976,wangComplexityguidedContainerReplacement2022},
run-time
analysis~\cite{kusumAdaptingGraphApplication2014,osterlundDynamicallyTransformingData2013,xuCoCoSoundAdaptive2013},
benchmarking~\cite{jungBrainyEffectiveSelection2011,coudercClassificationbasedStaticCollection2023},
or generating a set of collections that together implement a more
high-level description of program
state~\cite{hawkinsDataRepresentationSynthesis2011}. To our knowledge,
none of these are fully extensible, in the sense that a user or
library can add abstract data types, representations, operations, and
operation implementations, including, e.g., adding new representations
or operations to a previously defined abstract data type, without
modifying either the compiler or the original definition.

In this paper we present an approach that does support this degree of
extensibility, while still being capable of producing efficient
programs, allowing representation switching, and choosing
representations based on observable semantics, supported by a number
of compile-time analyses with various precision and efficiency
trade-offs.

Our approach centers around a single built-in \code{repr} type used
for \emph{all} types with compiler-picked representations. Each
abstract type defines an alias over some parameterization of
\code{repr}, and each representation provides a mapping from a
\code{repr} type to a concrete type. Operations are implemented as
overloadable functions, where each operation implementation may
specialize the type of the original operation by, e.g., picking
concrete representations for a \code{repr} type. Each operation
implementation has a cost annotation, and may also be implemented in
terms of other operations, in which case costs propagate.

Picking representations and operation implementations for a program is
thus a constrained optimization problem. We provide three kinds of
solvers: fast solvers with potentially inefficient solutions,
slow solvers with efficient solutions, and a solver that can transfer
solutions from a previous solution to a new one, even between modified
versions of the same program. This means that a user can use a fast
solver during iterative development, then run a slower solver to get a
better solution, and then retain the benefits of a better solution
during continued development as long as the program does not change
drastically.

We implement a compiler for a significant subset of OCaml and extend
it to use our approach for representation-flexible data structures. We
then use this compiler for two case studies. First, we implement a
universal collection type as a library: one type that represents all
conventional collections, e.g., sequences, sets, and maps. Second, we
implement a graph type with multiple backing representations as a library.

We also evaluate our implementation quantitatively through benchmarks
of handwritten as well as randomly generated programs, measuring both
compilation-time and run-time, as well as the optimality of solutions
produced by the heuristics-based solvers.

\noindent Concretely, our contributions are as follows:

\begin{itemize}
\item A fully extensible approach to specifying data structures whose
  representation is chosen automatically at compile-time
  (Section~\ref{sec:approach}).
\item A general approach for picking operation implementations across
  a program with multiple variants (Section~\ref{sec:implementation}).
\item Two libraries using our approach to provide a universal
  collection type and representation-flexible graphs, as qualitative
  case studies (Section~\ref{sec:case-studies}).
\item A compiler for a significant subset of OCaml, extended with
  \code{repr} types, along with a quantitative evaluation using
  benchmarks (Section~\ref{sec:evaluation}).
\end{itemize}

\sectionBreakMaybe

\section{Motivation and Overview}\label{sec:approach}

\begin{figure}[b]
  \centering
  \includegraphics{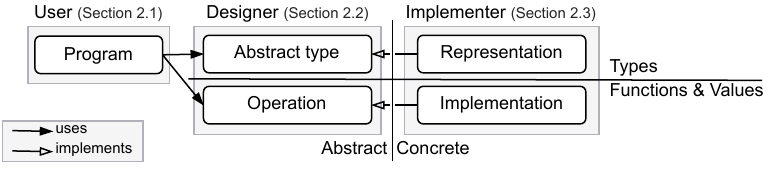}
  \caption{Overview of our approach split into three perspectives. A
    user writes a program that references abstract types and
    operations. An interface designer writes the library that defines
    these abstract types and operations. Finally, an implementer
    provides multiple concrete representations and implementations for
    each corresponding abstraction. Filled arrows denote ``uses'',
    while hollow arrows denote
    ``implements''.\label{fig:approach-overview}}
\end{figure}

In this section we give an overview of the design of our approach. As
a running example we consider a small library defining sequences with
two representations. A core component of our design is a
clear delineation between interface and implementation, thus we
explore our library from three perspectives: the user that programs
against the interface, the designer that specifies the interface, and
the implementer that provides concrete implementations underlying the
interface. These can be seen in
Fig.~\ref{fig:approach-overview}. Note that we use each of the terms
in Fig.~\ref{fig:approach-overview} with very specific intended
meanings throughout this paper. For example, an \emph{\impl} is always a
concrete implementation of an \emph{\operation}, while a \emph{\repr} is always a
concrete version of an \emph{\abstraction}. Note also that the example is designed to be relatively simple;
Section~\ref{sec:evaluation:case-uct} takes it significantly further
by expanding it to cover arbitrary collection types, not just
sequences.

Finally, we compare the components of our design with more conventional designs for abstract types in Section~\ref{sec:approach:others}.

\subsection{The User Perspective}\label{sec:approach:user}

\begin{figure*}
\begin{lstlisting}[language=RepCaml]
(* User-level code uses 'Seq' to represent a sequence without
 * considering the ideal choice of representation *)
let show_seq : ('a -> char seq) -> 'a seq -> char seq =
  fun f coll ->
    match split_first coll with
    | Some (head, tail) ->
      let intersperse a x = concat (concat a "; ") (f x) in
      let xs = foldl intersperse (f head) tail in $\label{line:seq-ex:foldl-use}$
      concat (concat "[" xs) "]"
    | None -> "[]"

let ex1 = show_seq show_int [1; 2; 3]
(* ex1 : char seq = "[1; 2; 3]" *)
let ex2 = show_seq (fun x -> x) ["a"; "b c"]
(* ex2 : char seq = "[a; b c]" *)
\end{lstlisting}
\caption{A function \code{show\_seq} for obtaining a pretty-printed
  version of a sequence, using the sequence library that we use as a
  running example for this section. The library introduces one type
  (\code{seq}) and several \operations (in this case \code{foldl},
  \code{split\_first}, and \code{concat}). Note that the compiler will
  automatically pick \reprs based on \operation usage. If we consider
  this function in isolation (ignoring places it might be used), the
  compiler will choose a cons-list for the input \code{seq} (which
  uses \code{split\_first} and \code{foldl}) and a rope for the output
  \code{seq} (which uses \code{concat}).\label{fig:seq-ex:user}}
\end{figure*}

\noindent Code using our example library is quite unremarkable on the
surface; all types and functions appear the same as normal types and
functions. However, some of these types and functions are
\abstractions and \operations, which are replaced by \reprs and \impls
during compilation. Fig.~\ref{fig:seq-ex:user} shows a function
producing a pretty-printed version of a sequence, with one
\abstraction (\code{seq}, parameterized by the type of its elements)
and three\footnote{Technically there are two additional
\operations. Our modified OCaml compiler does not change the
interpretation of list and string literals, thus these need to be
converted to \code{seq}s, which we accomplish with two \operations:
\code{seq\_of\_list} and \code{seq\_of\_string}. We omit them from
code examples in the interest of reducing clutter.} \operations
(\code{split\_first}, \code{concat}, and \code{foldl}). Notably, the
code makes no mention of underlying \reprs or \impls.

The choice of \repr is instead handled by the compiler and depends on
which \operations are used. For example, \code{show\_seq} uses three
operations: \code{split\_first}, \code{foldl}, and
\code{concat}. However, the compiler need not find one \repr with
efficient \impls of all three \operations. Instead, we note that the
\code{seq} consumed by \code{foldl} on
line~\ref{line:seq-ex:foldl-use} (which comes from the second
parameter of \code{show\_seq}) does not need to have the same \repr as
the \code{seq} it produces. This means that it is enough to have one
\repr with efficient \code{split\_first} and \code{foldl} \impls
(e.g., a cons-list), and another \repr with an efficient \code{concat}
\impl (e.g., a rope~\cite{boehmRopesAlternativeStrings1995}, which has
constant-time concatenation). We return to the details of how the
compiler knows of the efficiency of the various \impls in
Section~\ref{sec:approach:implementer}.

A consequence of this \repr-flexibility is that a user can freely
treat all sequences as though they had the same type (modulo
element-type, of course), i.e., all the same \operations are always
permissible, but without the performance penalty of needing one \repr
that supports all use cases.

\subsection{The Interface Designer Perspective}\label{sec:approach:designer}

The interface designer introduces new \abstractions and defines
\operations on them, forming the center of our approach: the user
programs against this interface, while the implementer provides the
underlying \reprs and \impls. Fig.~\ref{fig:seq-ex:designer} shows
an excerpt of the interface for the running example, and contains the
first explicit use of two of our new constructs.

\begin{figure*}
\begin{lstlisting}[language=RepCaml]
(* seq_t is a new (uninhabited) type *)
type 'a seq_t $\label{line:seq-ex:rseq-def}$
(* seq is a type alias, repr is a built-in type *)
type 'a seq = ('a seq_t) repr $\label{line:seq-ex:seq-alias}$

(* letop introduces an operation *)
letop prepend : 'a -> 'a seq -> 'a seq
letop append  : 'a seq -> 'a -> 'a seq
letop foldl   : ('acc -> 'a -> 'acc) -> 'acc -> 'a seq -> 'acc
letop foldr   : ('a -> 'acc -> 'acc) -> 'acc -> 'a seq -> 'acc

letop concat  : 'a seq -> 'a seq -> 'a seq
letop lookup  : 'k -> ('k * 'v) seq -> 'v option
\end{lstlisting}
\caption{The interface of our example library. The \code{repr} type is
  built-in and is used for all \abstractions whose \repr is to be
  decided by the compiler. Different \abstractions are distinguished
  by the type parameter passed to \code{repr}, which in this case is
  achieved by the newly defined type \code{seq\_t}. Operations are
  introduced by \code{letop} and have normal type signatures but no
  bodies; these are supplied later.\label{fig:seq-ex:designer}}
\end{figure*}

The first new construct is \code{repr}
(line~\ref{line:seq-ex:seq-alias}), which is a new built-in
type. \code{repr} has a single (phantom) type parameter, and is used
for \emph{all} \abstractions. We distinguish different \abstractions
by using different type arguments; in this example via the
uninhabited\footnote{We do not require types in this position to be
uninhabited, but since our system never uses values of this type they
tend to be.} type \code{seq\_t}, defined on
line~\ref{line:seq-ex:rseq-def}, which essentially acts as a
distinguishing tag. The type alias \code{seq} (on
line~\ref{line:seq-ex:seq-alias}) serves to hide this tag, making the
interface a bit more familiar.

The second new construct is \code{letop}, which declares a new
\operation. An \operation is essentially a \code{let}-declaration
without a body, and has the same static semantics: a variable
referencing an \operation with type \code{T} behaves the same as a
variable referencing a let-binding with type \code{T}. However, an
\operation has no inherent dynamic semantics; the compiler will
instead replace each reference to an \operation with a reference to
some concrete \impl. The \impls chosen must constitute a valid program
(e.g., an \impl expecting a rope must never be called with a
cons-list).

Next, we turn our attention to the implementer, who supplies \impls,
along with their expected \reprs and efficiency.

\subsection{The Implementer Perspective}\label{sec:approach:implementer}

The implementer provides the last two pieces to make our example
library usable: \reprs (concrete types to use in place of \code{repr}
types) and \impls (concrete functions to use in place of
\operations). An illustrative subset of these can be seen in
Fig.~\ref{fig:seq-ex:implementer}, which we now go through in more
detail.

\begin{figure*}
\begin{lstlisting}[language=RepCaml]
(* A representation is a mapping from (the type argument of) a Repr
 * type to some other type. *)
letrepr list_r {'a seq_t = 'a list}
letrepr rope_r {'a seq_t = 'a rope} $\label{line:seq-ex:rope}$
letrepr str_r {char seq_t = string}

(* The type of an implementation must be at least as specific as the
 * type of the operation, and it can assign representations via '!' *)
letimpl[1.0] prepend : 'a -> !list_r ('a seq) -> !list_r ('a seq) $\label{line:seq-ex:cons1}$
  = List.cons $\label{line:seq-ex:cons2}$
letimpl[n] prepend : char -> !str_r (char seq) -> !str_r (char seq) $\label{line:seq-ex:cons-str1}$
  = fun c str -> String.cat (String.make 1 c) str $\label{line:seq-ex:cons-str2}$
(* As a convenience we allow omitting parts of the type that are
 * unchanged from the operation *)
letimpl[n] append : !list_r _ -> _ -> !list_r _ $\label{line:seq-ex:snoc1}$
  = fun coll elem -> List.fold_right List.cons coll [elem] $\label{line:seq-ex:snoc2}$
letimpl[n] foldl : _ -> _ -> !list_r -> _ = List.fold_left $\label{line:seq-ex:foldl}$

(* An implementation can use other operations which allows for, e.g.,
 * default implementations *)
letimpl[1.0] concat = fun a b -> foldl append a b $\label{line:seq-ex:concat}$
(* These operations can be annotated with a cost scaling factor *)
letimpl[1.0] concat = fun a b -> foldr (@n prepend) b a $\label{line:seq-ex:concat2}$
\end{lstlisting}
\caption{A subset of the \reprs and \impls provided in our example
  library. A \repr (introduced by \code{repr}) specifies a possible
  way to replace a \code{repr} type with a concrete type, while an \impl
  (introduced by \code{letimpl}) does the same for \operations. Each
  \impl has a cost, an optional type signature, and a body. The body
  can contain any expression, including references to other
  \operations.\label{fig:seq-ex:implementer}}
\end{figure*}

\Reprs are introduced with \code{letrepr}, which describes a named
mapping from a \code{repr} type to a concrete type, expressed as a
transformation from the argument of the \code{repr}. The mapping may
contain type variables (e.g., \code{'a}) as stand-ins for types to be
preserved in the transformation. For example, suppose the compiler has
decided to use the \code{rope\_r}\footnote{In the actual library
representations typically reuse the name of the underlying type; in
the paper we use different names in the interest of clarity.} \repr
(defined on line~\ref{line:seq-ex:rope} in
Fig.~\ref{fig:seq-ex:implementer}) for a value of type \code{int
  seq}. Recall that \code{seq} is a type alias, thus \code{int seq} =
\code{(int seq\_t) repr}. The \code{rope\_r} definition matches
\code{'a seq\_t} (left-hand side on line~\ref{line:seq-ex:rope})
against \code{int seq\_t}, which binds \code{'a} to \code{int}, then
constructs \code{int list} (right-hand side on
line~\ref{line:seq-ex:rope}).

\Impls are introduced with \code{letimpl} by stating a cost, the
\operation to implement, a potentially specialized type, and a
body. The simplest, most explicit form of these are exemplified on
lines~\ref{line:seq-ex:cons1} and~\ref{line:seq-ex:cons2} (going
left-to-right):

\begin{itemize}
\item The cost of an \impl is given in square brackets and is an
  expression that must evaluate to a floating point number, in this
  case merely the literal \code{1.0}.
\item The \operation to be implemented is \code{prepend}.
\item The type of an \impl may be a more specialized version of the
  type of the \operation, either by being less polymorphic, or by
  specifying concrete \reprs. An exclamation mark is used to assign
  \reprs: \code{!r t} means that the type \code{t} (which must be a
  \code{repr} type) has the \repr \code{r}. In this case the \impl is
  still polymorphic in the element type, but requires that both
  \code{seq}s have the \code{list\_r} \repr.
\item Finally, the body can be found on the right-hand side of the
  equals sign, on line~\ref{line:seq-ex:cons2}. The body must be
  typeable under the annotated signature, \emph{after} substituting
  each \code{repr} type via its explicit \repr, if any. In this case,
  the expression \code{List.cons} must have type \code{'a -> 'a list
    -> 'a list}, which is indeed the case.
\end{itemize}

\noindent Next, we consider each of the components of a \code{letimpl}
in more detail.

A cost annotation is an arbitrary expression which, most importantly,
can use certain free variables. For example, the definitions on
lines~\ref{line:seq-ex:cons-str1}, \ref{line:seq-ex:snoc1},
and~\ref{line:seq-ex:foldl} use the variable \code{n}. Such variables
have no pre-defined or intrinsic meaning to our system; they are given
as constants by the user at compile-time, presently via command-line
flags. Our example library uses \code{n} to represent the size of a
sequence, which---given a suitably large \code{n}---approximates
algorithmic time complexity. When the compiler chooses \impls it tries
to minimize the sum of their costs.

The annotated type will typically differ from that of the
corresponding \operation (e.g., line~\ref{line:seq-ex:cons1} uses the
\code{list\_r} \repr, while line~\ref{line:seq-ex:cons-str1} uses
\code{str\_r} and additionally constrains the element type to be
\code{char}). However, the difference is often slight, thus we allow
replacing the unchanged parts with an underscore, as a
convenience. For example, the \impl on line~\ref{line:seq-ex:snoc1}
uses \code{\_} in place of each of its parameters and the return type,
specifying only its \reprs. Finally, we allow \code{!r} as syntactic
sugar for \code{!r \_} (used on line~\ref{line:seq-ex:foldl}), and
interpret an omitted type annotation as an annotation consisting of a
single underscore, i.e., the type is precisely the same as for the
\operation (e.g., lines~\ref{line:seq-ex:concat}
and~\ref{line:seq-ex:concat2}).

The body of an \impl can be any arbitrary expression, but typically
takes one of two forms: a reference to a previously implemented
function, or a default \impl written in terms of other
\operations. When an \impl uses other \operations the compiler must
recursively choose corresponding \impls, and the final cost becomes a
sum of all chosen \impls. For example, the \code{concat} \impl on
line~\ref{line:seq-ex:concat} needs an \impl for \code{foldl} and
\code{append} as well, and the final cost becomes $\code{1.0} +
c_{\code{foldl}} + c_{\code{append}}$, where $c_{\code{foldl}}$ and
$c_{\code{append}}$ are the costs of the chosen \impls\footnote{We
give each default \impl in our example library an annotated cost of
(at least) \code{1.0}, ensuring that they have a higher cost than the
sum of their parts, which gently guides the compiler towards
specialized \impls that do not have this extra cost.}. Of course, this
is a poor approximation of the actual algorithmic complexity of the
\impl; \code{append} will be called \code{n} times, thus the final
term should be $\code{n}\cdot c_{\code{append}}$, not
$c_{\code{append}}$. We can achieve this with an additional explicit
annotation; \code{@c op} is the same as \code{op}, except its cost is
multiplied by \code{c}. For example, the next \impl
(line~\ref{line:seq-ex:concat2}) annotates its use of \code{prepend} to get a more
accurate final cost of $\code{1.0} + c_{\code{foldr}}
+ \code{n} \cdot c_{\code{prepend}}$.

\subsection{Closely Related Pre-Existing Concepts}\label{sec:approach:others}

This section gives a high-level comparison of our approach and the
most closely related concepts in common use, and why the latter are
insufficient for our intended use cases.

\emph{Typeclasses}~\cite{wadlerHowMakeAdhoc1989} are notably similar
to how we handle our operations. In particular, we can mostly view
each \code{letop} as a single method typeclass and each \code{letimpl}
as an instance. However, the two approaches solve subtly different
problems. Instance resolution for typeclasses starts with an
invocation and mostly concrete types, then selects an appropriate
instance. In contrast, we start with an invocation and unknown
\code{repr} types, then select implementations \emph{and \reprs}. This
means that our design must be able to select types as part of
implementation resolution, which is typically undesirable for instance
resolution. For example, the Haskell compiler GHC might note that a
type variable is ambiguous and list potential instances in the type
error, but not pick any of them. A notable consequence is that
instance resolution can be done independently for each invocation,
while finding implementations by necessity cannot; \opUses might need
to share \reprs.

Note also that typeclasses as used in, e.g., Haskell require
\emph{coherence}, i.e., that each type should have at most one
possible instance. In contrast, multiple \impls per \operation are the
norm in our system, largely due to default implementations.

\emph{Implicit
parameters}~\cite{oderskySimplicitlyFoundationsApplications2017,oliveiraImplicitCalculusNew2012}
can be used in a similar fashion to typeclasses, and they do not have
coherence requirements. However, types are still considered to be
inputs to the resolution mechanism, i.e., they could not be used to
select \reprs.

\emph{Parametric polymorphism} could conceivably be used to fill a
similar role to our \code{repr} types. In theory, we could add a
quantified type variable per unknown representation and pass
operations as extra arguments, then solve for efficient
representations and operations and add those outside the
program. However, universally quantified type variables are not an
ideal fit in this context because of their precision. In particular,
each instantiation must replace each type variable with precisely one
type, and distinct type variables are considered to be distinct
types. This means that any explicit type signatures in the program
must explicitly note which representations must be equal and which
ones may differ, which is tedious and error-prone to update
manually. In contrast, a programmer can treat two \code{repr} types as
the same type (assuming equal type parameters), even though they may
be given distinct \reprs later on.

An alternate view of our approach is that each \code{repr} type hides
a type variable, and our \repr analysis
(cf. Section~\ref{sec:repr-analysis}) determines which of these must
be equal. Our solvers then replace each type variable with a concrete
\repr, though this is done \emph{inside} the program, rather than
outside, in contrast with the imagined approach of the previous
paragraph.

In an \emph{object-oriented} view we typically have, e.g., an
inheritance hierarchy of interfaces for different collections like
\code{Collection}, \code{List}, and \code{Set}. However, a subtype
must implement all \operations of its supertype, which conflicts with
supplying efficient specialized \reprs that support only a subset of
these \operations. Notably, the apparent interface available to the
user in our approach includes \emph{all} \operations, while the
selected concrete type might only support the subset actually
used. Much like for typeclasses we can achieve a similar result by
using one single-method interface per \operation, then substituting an
appropriate combined interface supporting only the \operations in use
instead of the full abstract type during \repr analysis.

\sectionBreakMaybe

\section{Design and Approach}\label{sec:implementation}

\begin{figure*}
  {\centering
    \includegraphics{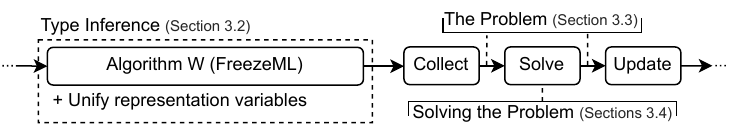}
  \par}

  \caption{An overview of the relevant parts of our
    implementation. First, we extend a unification-based type
    inference algorithm (in our case the already extended version of
    Algorithm W used in
    FreezeML~\cite{emrichFreezeMLCompleteEasy2020}) with
    \emph{representation variables} to track which \code{repr} types
    must have the same representation. Second, we collect all
    operation uses across the program, solve a constrained
    optimization problem to determine concrete implementations for all
    of them, then update the program to use the chosen
    implementations. Note that the last step removes our constructs
    from the program; they are all replaced with normal let-bindings
    and variables.\label{fig:impl-overview}}
\end{figure*}

This section describes the intuition and implementation strategy of
our approach and how we pick \impls. An overview can be found in
Fig.~\ref{fig:impl-overview}. Broadly speaking we implement our
approach by lowering from a program containing \operations and
\code{repr} types to a program that does not, i.e., an ordinary OCaml
program (example in
Section~\ref{sec:intuition-running-example}). First, we use a modestly
extended type inference algorithm (described in
Section~\ref{sec:repr-analysis}) to determine which \code{repr} types
must share \reprs, then we collect all \operation uses, find \impls
for them, and update the program. Finding valid \impls is not a
straight-forward process and is, in general, not always possible; we
formulate it as a constrained optimization problem in
Section~\ref{sec:optimization-problem}, then describe our solving
approach in Section~\ref{sec:solving-the-problem}.

Note also that from this section on we refer to the call-site of an
\operation as an \emph{\opUse}, i.e., each occurrence of an \operation
except its declaration (\code{letop}) or definition (\code{letimpl})
is an \opUse.

\subsection{Intuition and Running Example}\label{sec:intuition-running-example}

\begin{figure*}
\begin{lstlisting}[language=RepCaml]
letrepr list_r {'a = 'a list} $\label{line:implementation:running-example:repr1}$
letrepr iset_r {int = IntSet.t} $\label{line:implementation:running-example:repr2}$
letop fold : ('a -> 'b -> 'a) -> 'a -> 'b repr -> 'a $\label{line:implementation:running-example:start-ops}$
letimpl[n] fold : _ -> _ -> !list_r (_ repr) -> _ = List.fold_left
letimpl[n] fold : _ -> _ -> !iset_r (_ repr) -> _ = IntSet.fold
letop contains : 'a -> 'a repr -> bool
letimpl[min n W] contains : _ -> !iset_r (_ repr) -> _ = IntSet.contains
letimpl[n] contains = fun elem coll -> $\label{line:implementation:running-example:default1}$
  fold (fun x found -> found || x = elem) false coll $\label{line:implementation:running-example:default2}$ $\label{line:implementation:running-example:end-ops}$
let hasTwo coll = contains 2 coll $\label{line:implementation:running-example:contains-use}$
\end{lstlisting}
\hspace{0.28\textwidth}\hfill{}\hphantom{or}\hfill\begin{minipage}{0.30\textwidth}
  \begin{lstlisting}[language=RepCaml,gobble=4]
    let fold =
      List.fold_left
  \end{lstlisting}
\end{minipage}\hfill{}or\hfill
\begin{minipage}{0.30\textwidth}\vspace{0pt}%
  \begin{lstlisting}[language=RepCaml,gobble=4]
    let fold =
      IntSet.fold
  \end{lstlisting}
\end{minipage}\\[-7pt]
\begin{minipage}{0.28\textwidth}\vspace{0pt}%
  \begin{lstlisting}[language=RepCaml,gobble=4,firstnumber=3]

    let contains =
      IntSet.contains
  \end{lstlisting}
\end{minipage}\hfill{}or\hfill
\begin{minipage}{0.66\textwidth}\vspace{0pt}%
  \begin{lstlisting}[language=RepCaml,gobble=4,firstnumber=3]
    let contains = fun elem coll ->
      fold (fun x found -> found || x = elem)
        false coll
  \end{lstlisting}
\end{minipage}\\[-7pt]
\begin{lstlisting}[language=RepCaml,gobble=2,firstnumber=6]
  let hasTwo coll = contains 2 coll
\end{lstlisting}
\caption{The running example for Section~\ref{sec:implementation}
  (above) and its three possible transformed versions (below). For brevity we use \code{'a repr} to mean a set of
  \code{'a}, rather than something like \code{('a uset) repr}, as we
  suggest in Section~\ref{sec:approach:designer}. The example contains
  two \reprs: \code{list\_r} for cons-lists and \code{iset\_r} for an
  optimized set implementation that only handles \code{int}
  values. The first three \impls reference an external function while
  the last is a default implementation written in terms of another
  operation. The example has a single \opUse outside a \code{letimpl}
  (on the last line), for which our implementation must pick an
  \impl.\label{fig:implementation:running-example}}
\end{figure*}

\noindent This section uses the upper half of
Fig.~\ref{fig:implementation:running-example} as a small running
example. The example defines a predicate \code{hasTwo} that is true if
a given collection contains \code{2}, using \code{repr} types as
collections. In the interest of brevity we here use \code{'a repr} to
mean a set of \code{'a}s, rather than something like \code{('a uset)
  repr} that we advocate in Section~\ref{sec:approach:designer}. In
total, the example has two \reprs (cons-lists and int-sets\footnote{We
use patricia trees~\cite{morrisonPATRICIAPracticalAlgorithm1968},
whereby the \code{contains} \impl has cost \code{min(n, W)}, where
\code{W} is the number of bits in an \code{int} on the system.}, first
two lines), two \operations (with two \impls each,
lines~\ref{line:implementation:running-example:start-ops}-\ref{line:implementation:running-example:end-ops}),
and two \opUses (\code{fold} on
line~\ref{line:implementation:running-example:default2} and
\code{contains} on
line~\ref{line:implementation:running-example:contains-use}). The
final \impl (lines~\ref{line:implementation:running-example:default1}
and~\ref{line:implementation:running-example:default2}) is implemented
in terms of another \operation, namely \code{fold}, while the others
merely refer to imported functions.

Intuitively there are three valid ways to transform the program:

\begin{enumerate}
\item\label{item:iset-contains} Resolve the use of \code{contains} to
  \code{IntSet.contains} (lower left of
  Fig.~\ref{fig:implementation:running-example}).
\item\label{item:default-list} Resolve \code{contains} to the default
  implementation using \code{List.fold\_left} (lower middle of
  Fig.~\ref{fig:implementation:running-example}).
\item\label{item:default-iset} Resolve \code{contains} to the default
  implementation using \code{IntSet.fold} (lower right of
  Fig.~\ref{fig:implementation:running-example}).
\end{enumerate}

\noindent Of these, version~\ref{item:iset-contains} has the lowest
cost, while version~\ref{item:default-list} has the most general
type. Version~\ref{item:default-iset} is redundant; it is only a valid
choice when the element type is \code{int}, in which case
version~\ref{item:iset-contains} is also valid and has a lower cost.

\subsection{Type Inference \& Representation Analysis}\label{sec:repr-analysis}

The first step of finding \impls for \opUses is to determine the
relevant constraints: primarily which \code{repr} types must have the
same \repr, but also which \impls have less general types than the
\operations they implement. Both kinds of constraint affect what \impl
choices produce a well-typed program, thus both must be taken into
account. Of course, the type of an \impl is explicitly given in the
program, i.e., it is easy to find, thus we turn our attention to how
we determine if two \code{repr} types must have the same \repr.

The key observation underlying our approach here is that a
unification-based type inference algorithm\footnote{We extend the
modified version of Algorithm
W~\cite{damasPrincipalTypeschemesFunctional1982} presented by Emrich
et al.~\cite{emrichFreezeMLCompleteEasy2020} for FreezeML, but the
core idea is applicable to any unification-based algorithm.} attempts
to unify two types when values of one type need to be used in a place
expecting a value of the other type, i.e., it is a conservative
data-flow analysis. We exploit this by adding a \emph{\reprVar} to
each \code{repr} type that functions much the same as a unification
variable; when two \code{repr} types are unified we also unify their
\reprVars. New \reprVars are generated in two cases:

\begin{enumerate}
\item\label{item:annot-repr-var} Each \code{repr} type occurring in an
  explicitly given type in the program (e.g., in type annotations or
  type definitions) is given a new \reprVar (\emph{after} expanding
  type aliases, if present).
\item\label{item:instantiate-op-use} When instantiating the type of an
  \operation in an \opUse we create new \reprVars for each \code{repr}
  type in the result.
\end{enumerate}

\noindent For example, consider the code in
Fig.~\ref{fig:implementation:running-example}, reproduced below with
additional annotations to show the result of running type inference
with our \repr analysis. We use distinct subscripts to denote distinct
\reprVars, and put the subscripts on explicit \code{repr} types as
well as variables and \opUses whose inferred type contains a
\code{repr} type. Note that the \code{repr} types in explicit type
signatures are given distinct \reprVars, following
point~\ref{item:annot-repr-var} above. Note also that the \reprVar in
the definition of \code{fold} is distinct from its use on
line~\ref{line:implementation:repr-vars:fold-use}. This is because the
\opUse gets a fresh \reprVar, following
point~\ref{item:instantiate-op-use} above.

\begin{lstlisting}[language=RepCaml,firstnumber=3]
letop fold : ('a -> 'b -> 'a) -> 'a -> 'b repr$\ensuremath{_1}$ -> 'a $\label{line:implementation:repr-vars:fold-op}$
letimpl[n] fold : _ -> _ -> !list_r (_ repr$\ensuremath{_2}$) -> _ = List.fold_left
letimpl[n] fold : _ -> _ -> !iset_r (_ repr$\ensuremath{_3}$) -> _ = IntSet.fold
letop contains : 'a -> 'a repr$\ensuremath{_4}$ -> bool
letimpl[min n W] contains : _ -> !iset_r (_ repr$\ensuremath{_5}$) -> _ = IntSet.contains
letimpl[n] contains : _ -> _ repr$\ensuremath{_6}$ -> _ = fun elem coll ->
  fold$\ensuremath{_6}$ (fun x found -> found || x = elem) false coll$\ensuremath{_6}$ $\label{line:implementation:repr-vars:fold-use}$
let hasTwo coll = contains$\ensuremath{_7}$ 2 coll$\ensuremath{_7}$
\end{lstlisting}

\noindent In contrast, a hypothetical later call to \code{hasTwo}
would not generate a fresh \reprVar, it would re-use \code{repr}$_7$, since
\code{hasTwo} is a normal \code{let}-binding, not a \code{letop}.

The sharing between uses of \code{let}-bound values can occasionally
be inconvenient, e.g., if a programmer chose to use a
\code{let}-binding for a function that ends up being used on many
otherwise independent \code{repr} types, since it may significantly
constrain \repr choices. For this reason our compiler uses a slight
extension to the core approach: non-recursive \code{let}-bindings
whose right-hand side is a value (e.g., a function) and whose type
contains a \code{repr} type are replaced with a \code{letop} and
\code{letimpl} pair.

\subsection{The Constrained Optimization Problem}\label{sec:optimization-problem}

\begin{figure*}
  \begin{tabular}{@{}l@{ }l@{ }l@{}}
    $\mOpUse$ & $::=$ & $\mCost \times \mOp \times \mType$ \\
    $\mImpl$ & $::=$ & $\mCost \times \mOp \times \mType \times \many{\mOpUse}$ \\
    $\mSol$ & $::=$ & $\mImpl \times \many{\mSol}$ \\
    \addlinespace
  \end{tabular} \qquad
  \begin{tabular}{@{}r@{ }r@{ }l@{\qquad}r@{}}
    $\cost{\mSol}$ & $:=$ & $\projectTwo{\mSol}{\mImpl}{\mCost}$ & $\boxed{\costBare : \mSol \rightarrow \mCost}$ \\
    & $+$ & \multicolumn{2}{@{}l@{}}{$\displaystyle\sum_i \cost{\getAt{\project{\mSol}{\many{\mSol}}}{i}}\cdot \project{\getAt{\project{\project{\mSol}{\mImpl}}{\many{\mOpUse}}}{i}}{\mCost}$} \\
  \end{tabular}
  \caption{The tuples for \opUses, \impls, and \solutions that our
    solvers interact with (left), and the cost of a \solution
    (right). We use $\many{x}$ to denote a sequence of $x$,
    $\getAt{x}{i}$ to denote the $i$th element of sequence $x$, and
    $\project{x}{\mCost}$ (and $\project{x}{\mImpl}$,
    $\project{x}{\many{\mOpUse}}$, etc.) to denote the $\mCost$
    (resp. $\mImpl$, $\many{\mOpUse}$, etc.) element of the tuple
    $x$.\label{fig:solver-types-and-cost}}
\end{figure*}

The analysis required to pick efficient \impls for a given program is,
in general, a global analysis. This is because \opUses may refer to
\code{repr} types that share \reprVars, meaning that an efficient
choice for one \opUse may constrain another to a suboptimal choice,
leading to a higher cost overall. As such, each of our solvers receive
as input all \opUses in a program, each paired with the \impls in
scope at that point in the program.
Fig.~\ref{fig:solver-types-and-cost} shows the components of these
on the left. An $\mOpUse$ is a tuple containing a scaling factor, the
\operation used, and the inferred type of the \opUse. Similarly, an
$\mImpl$ is a tuple containing a cost, the \operation implemented, the
annotated type (after filling in omitted types), and the \opUses
within the body of the corresponding \code{letimpl}.

For example, the running example
(Fig.~\ref{fig:implementation:running-example}) contains four \impls
(using subscripts to note attached \reprVars, as in the previous
section, and renaming type variables to ensure them all unique):

\begin{itemize}
\item $\mathit{fold} := (\code{n}, \code{fold}, \code{('a -> 'b -> 'a) -> 'a -> !list\_r ('b repr$_2$) -> 'a}, [])$
\item $\mathit{fold}' := (\code{n}, \code{fold}, \code{('c -> int -> 'c) -> 'c -> !iset\_r (int repr$_3$) -> 'c}, [])$
\item $\mathit{contains} := (\code{min n W}, \code{contains}, \code{int -> int repr$_5$ -> bool}, [])$
\item $\mathit{contains}' := (\code{n}, \code{contains}, \code{'d -> 'd repr$_6$ -> bool}, [\mathit{use}])$, where\\
  $\mathit{use} := (\code{1.0}, \code{fold}, \code{(bool -> 'd -> bool) -> bool -> 'd repr$_6$ -> bool})$
\end{itemize}

\noindent Note that the \opUse inside $\mathit{contains}'$ has the
default scaling factor of \code{1.0}, and that \code{'d} is the
\code{'d} bound by the type signature of $\mathit{contains}'$, rather
than a new type variable. The running example has only one \opUse
outside a \code{letimpl}, thus the input of the constrained
optimization problem for the running example is a list containing a
single pair:

\[
([\mathit{fold}, \mathit{fold}', \mathit{contains}, \mathit{contains}'], (\code{1.0}, \code{contains}, \code{int -> int repr$_7$ -> bool}))
\]

\noindent A \emph{\solution} is then one $\mImpl$ per $\mOpUse$ in the
input, including recursively, i.e., a tree of \impls. Formally we
denote such a tree rooted at a single \impl by $\mSol$, which is a
pair consisting of an $\mImpl$ and a sequence with one $\mSol$ for
each $\mOpUse$ in that $\mImpl$. Of course, not any \solution will do,
we wish to pick the lowest cost \solution that is valid.

The cost of a \solution is defined by the recursive $\costBare$
function, found to the right in
Fig.~\ref{fig:solver-types-and-cost}, as the sum of the cost of the
chosen \impl and the scaled cost of each of its sub-\solutions.

Intuitively speaking, a \solution is \emph{valid} if the resulting
program type-checks and each \reprVar is assigned a single \repr. For
example, $[(\mathit{contains}', [(\mathit{fold}, [])])]$ is a valid
\solution because:

\begin{itemize}
\item The type of $\mathit{contains}'$ can be unified with $\code{int
  -> int repr$_7$ -> bool}$, and the type of $\mathit{fold}$ can be
  unified with the type of $\mathit{use}$.
\item \ReprVar 2 (and variables 6 and 7, through unification) is
  assigned the \code{list\_r} \repr.
\end{itemize}

\noindent For a formal description of validity, including more
complicated cases such as handling \reprVars occurring outside
\code{letimpl}s and using the same \impl multiple times in a
\solution, see the supplementary material.

Next we turn our attention to producing these \solutions
automatically.

\subsection{Solving the Problem}\label{sec:solving-the-problem}

Our implementation contains several solvers with various trade-offs,
primarily between completeness (always finding an optimal solution if
one exists) and speed. There is significant overlap between these
solvers, primarily in how they structure their input. Conceptually,
the problem to solve is a decision tree of conjunctions and
disjunctions: for each \opUse we must chose one of several \impls
(disjunction) and for each \impl there are zero or more \opUses to
satisfy (conjunction). Each solver traverses and transforms this tree
in different ways.

We have one \code{exhaustive} solver that produces all possible
\solutions, mostly used as a debugging tool and sanity check, since it does not scale beyond very small examples. It works by traversing the tree bottom-up, transforming it
into disjunctive normal form along the way.
The remaining solvers, 3 complete and 2 heuristics-based, produce a
single \solution. They all start by partitioning the input program
into independent sub-problems; subtrees that share no unification or
\reprVars and can thus be solved independently, and then use different
approaches:

\begin{description}
\item[\code{bottom-up}] works like \code{exhaustive}, except it prunes
  redundant \solutions at each step. A \solution is redundant if it is
  neither cheaper nor more general than another \solution. Note that
  the generality check ignores details that are internal to each
  \solution, e.g., \reprVars that only occur inside that \solution and
  no other.
\item[\code{greedy}] recursively creates a lazy list of non-redundant
  \solutions at each level of the tree in cheapest-first order, and
  then tries to retrieve the first element of the top-most list.
\item[\code{z3}] encodes the tree as an SMTLIB2 model and solves it
  using Z3~\cite{demouraZ3EfficientSMT2008a}.
\item[\code{mixed}] uses \code{bottom-up} if the input is small
  (\solution space $\leq 100000$) and otherwise falls back to heuristics. There are two heuristics: (i) trying
  to assign the same \repr to as many \reprVars as possible, and (ii) repeatedly picking the cheapest option for each child, picking those that are consistent with each other (e.g., same \repr assignments), and finding new options for the other children while only considering those consistent with the already picked children. Note that
  both of these may provide several \solutions, in which case we pick the cheapest.
\item[\code{transfer}] takes a \solution to a previous problem
  (typically from a previous compilation) and tries to apply it to the
  current problem. Each sub-problem is matched against each
  sub-\solution, leaving only those that occur somewhere in a previous
  \solution. This forms a sort of filter over the tree, after which we
  apply \code{mixed} to pick between the (typically quite few,
  relatively speaking) remaining alternatives.
\end{description}

\noindent For more details on the workings of each of these, see the supplementary material. Next, we turn our attention to the task of evaluating our
approach in general, as well as the performance and results of our
solvers when applied to concrete programs.

\sectionBreakMaybe

\section{Case Studies and Qualitative Evaluation}\label{sec:case-studies}

We evaluate the qualitative aspects of our approach by implementing
two libraries; one providing a universal collection type, and one
providing a representation-flexible graph. We make three claims
relevant to this section:

\begin{claims}
\item\label{claim:extensible} Our approach permits fully extensible
  libraries, i.e., a user can add new  \abstractions, \reprs, \operations, and \impls without modifying either the compiler or the original
  library.
\item\label{claim:repr-switching} Libraries using \code{repr} types
  support \repr switching.
\item\label{claim:selection-flexibility} Libraries using \code{repr}
  types can be designed to allow \repr selection based on observable
  semantics, as opposed to requiring equivalent semantics for
  \emph{all} possible \operations.
\end{claims}

\noindent We begin by describing each of the two libraries, then cover
each of the claims in turn.

\subsection{A Universal Collection Type as a Library}\label{sec:evaluation:case-uct}

The first library implements a universal collection type (UCT) using
\code{repr} types. By ``universal'' we mean that it covers all
homogenous collections (i.e., all elements in a collection have the
same type) that can grow, shrink, and be queried. This excludes, e.g.,
matrices and vectors (in the mathematical sense) as these are fixed in
size. There are three additional key aspects of our design, beside the
universality mentioned above:

\begin{itemize}
\item The user should be able to create a collection with no specified iteration order, thereby giving the compiler greater freedom in
  finding an efficient \solution. This is primarily relevant for,
  e.g., sets, which have no semantic element order, yet a program
  iterating over a set will observe some order.
\item There should be at least one \repr that supports all \operations
  (i.e., all programs have a \solution).
\item There should be a clear definition of what it means for an \impl
  to be semantically correct\footnote{An interesting direction for
  future work is to check correctness automatically, e.g., through
  property-based testing in the vein of QuickSpec~\cite{smallboneQuickSpecificationsBusy2017}.}.
\end{itemize}

\noindent The core intuition for our design is as follows: a
collection is a sequence of elements, containing a subset of inserted
elements (e.g., a sequence retains all elements, while a set removes
duplicates) in some order (e.g., a sequence maintains insertion order,
while a sorted set maintains a sorted order). We encode this by
defining a type alias over a \code{repr} type, as done in
Section~\ref{sec:approach:designer}, with two type parameters: the
element type, and a pair of properties. Fig.~\ref{fig:uct-types}
shows the core type definitions\footnote{The property types
(\code{keep\_*} and \code{order\_*}) are technically not fundamental,
in the sense that properties can be defined anywhere, but these
properties in particular are common enough that they warrant inclusion
in the core library.} (left) and type aliases for some familiar kinds
of collections (right).

\begin{figure}
  \begin{minipage}{0.37\textwidth}
    \begin{lstlisting}[language=RepCaml,gobble=6,name=uct-def]
      type ('a, 'p) ucoll

      type keep_all
      type keep_last
      type keep_last_key

      type order_seq
      type order_sorted
      type order_sorted_key

      type ('a, 'p) coll =
        (('a, 'p) ucoll) repr
    \end{lstlisting}
  \end{minipage}\hfill
  \begin{minipage}{0.61\textwidth}
    \begin{lstlisting}[language=RepCaml,gobble=6,name=uct-def,firstnumber=auto]
      type 'a seq =
        ('a, keep_all * order_seq) coll

      type 'a set =
        ('a, keep_last * _) coll
      type 'a sorted_set =
        ('a, keep_last * order_sorted) coll

      type ('k,'v) map =
        ('k*'v, keep_last_key * _) coll
      type ('k,'v) ordered_map =
        ('k*'v, keep_last_key * order_seq) coll
    \end{lstlisting}
  \end{minipage}
  \caption{The central type definitions for our universal collection
    type. On the left are the basic definitions: \code{ucoll} is the
    tag to distinguish universal collections from other \code{repr}
    types, \code{keep\_*} and \code{order\_*} are properties, and
    \code{coll} is the central type alias. On the right are type
    aliases for some common kinds of collections: \code{seq} for
    sequences, \code{set} and \code{sorted\_set} for sets
    (compiler-picked element order and sorted order, respectively),
    and \code{map} and \code{ordered\_map} for maps (compiler-picked
    element order and insertion order,
    respectively).\label{fig:uct-types}}
\end{figure}

Note that the \code{ucoll}, \code{keep\_*}, and \code{order\_*} types
are ordinary types (though uninhabited); their semantic meaning stems
entirely from \impls using them consistently as tags for their
semantics. Of course, this means that any programmer or library can
add additional properties for more exotic kinds of collections, which
gives us universality. Note also that we encode a map as a collection
of key-value pairs, with properties that specifically consider the key
(e.g., \code{keep\_last\_key} and \code{order\_sorted\_key}). Finally,
note that we express order-independence using the special \code{\_}
type. This type matches with any other type during \impl selection,
e.g., a collection with properties \code{keep\_last * \_} could match
an \impl requiring properties \code{keep\_last * order\_sorted}.

Moving on to \operations, we divide them into two separate groups:
fundamental \operations and composite \operations. The fundamental
\operations are a fixed set of five \operations (\code{empty},
\code{append}, \code{prepend}, \code{foldl}, and \code{foldr}), while
all other \operations are composite. Composite \operations are
required to have default \impls in terms of fundamental \operations,
i.e., a \repr that supports all five fundamental \operations has at
least one \impl for \emph{every} \operation, with the caveat that some
\impls may be quite inefficient. Note that while the library considers
fundamental \operations different from composite \operations, the
compiler sees no such distinction; it is simply a convention we follow
to ensure that no program is without \solution.

For example, the code below shows the declarations of three of the
fundamental \operations and two composite \operations, along with
default \impls for the latter.

\begin{lstlisting}[language=RepCaml,gobble=2]
  letop empty : ('a, 'p) coll
  letop foldl : ('a -> 'b -> 'a) -> 'a -> ('b, 'p) coll -> 'a
  letop prepend : 'a -> ('a, 'p) coll -> ('a, 'p) coll

  letop map : ('a -> 'b) -> ('a, 'p) coll -> ('b, 'p) coll
  letimpl[1.0] map = fun f xs ->
    foldl (fun acc x -> @n append acc (f x)) empty xs

  letop concat : ('a, 'p) coll -> ('a, 'p) coll -> ('a, 'p) coll
  letimpl[1.0] concat = fun a b ->
    foldl (fun acc x -> @n append acc x) a b
\end{lstlisting}

\noindent Of course, since the fundamental \operations are symmetric
we can provide two additional default \impls by replacing \code{foldl}
with \code{foldr} and \code{append} with \code{prepend}. We omit these
above for brevity, but both are present in the library.

For each \repr in the library we give \code{letimpl}s for the
fundamental \operations they support (they do not need to support all
of them), along with further \code{letimpl}s for composite \operations
for which they support a more efficient \impl than the default.

As for correctness, the semantics of each composite \operation is
defined as the observable behavior of its default \impl. Equivalently,
two \impls that are valid for an \opUse with a fully concrete inferred
type (i.e., no underscores) should be observably indistinguishable.

Concretely, the library contains 10 \reprs:

\begin{itemize}
\item \code{list} and \code{snoc}, both of which use OCaml's built-in
  list type, but store elements in different orders, left-to-right or
  in reverse, respectively.
\item \code{string}, which uses OCaml's built-in string type.
\item \code{rope} and \code{str\_rope} using
  ropes~\cite{boehmRopesAlternativeStrings1995}, where
  \code{str\_rope} is specialized to character sequences.
\item \code{rbtree-set} and \code{rbtree-map} using red-black
  trees~\cite{guibasDichromaticFrameworkBalanced1978} to implement
  sets and maps. The implementation follows Cornell's CS3110 text book
  (\url{https://cs3110.github.io/textbook}).
\item \code{avl-set} and \code{avl-map} using AVL
  trees~\cite{adelson-velskiiAlgorithmOrganizationInformation1962},
  ported from OCaml's standard library to use compare functions via
  \code{letop}s instead of functors, to implement sets and maps.
\item \code{binom-queue}, a binomial queue.
\end{itemize}

\subsection{A Representation-Flexible Graph Type as a Library}\label{sec:evaluation:case-ugt}

As a second case study and further exploration of the expressiveness
of our approach we have implemented a library for directed graphs with
weighted edges. The interface of a graph has fundamental operations
for adding, removing and querying the presence of vertices and
edges. Just like the universal collection types of Section~\ref{sec:evaluation:case-uct}
these fundamental operations give us default implementations of
composite operations such as building a graph from a list of vertices and edges and
querying all the neighbors of a vertex.
A difference from universal collection types is that the cost
annotations now contain two variables: one representing the number
of vertices \code{v} and one representing the number of edges
\code{e}.
Fig.~\ref{fig:graph-types} shows the full interface of our graph
library.

\begin{figure}
\begin{lstlisting}[language=RepCaml,gobble=2]
  type ('v, 'e) ugraph
  type ('v, 'e) graph = (('v, 'e) ugraph) repr

  (* Fundamental operations (excerpt) *)
  letop empty_graph : ('v, 'e) graph
  letop add_vertex : ('v, 'e) graph -> 'v -> ('v, 'e) graph
  letop add_edge : ('v, 'e) graph -> 'v -> 'e -> 'v -> ('v, 'e) graph
  letop has_vertex : ('v, 'e) graph -> 'v -> bool
  letop has_edge : ('v, 'e) graph -> 'v -> 'v -> bool
  letop outgoing_edges : ('v, 'e) graph -> 'v -> ('e * 'v) list

  (* Composite operations *)
  letop neighbors : ('v, 'e) graph -> 'v -> 'v list
  letimpl[1.0] neighbors = fun g v ->
    List.map Stdlib.snd (outgoing_edges g v)

  letop new_graph : 'v list -> ('v * 'e * 'v) list -> ('v, 'e) graph
  letimpl[1.0] new_graph = fun vs es ->
    List.fold_right (fun (v1, e, v2) g -> @e add_edge g v1 e v2) es
    (List.fold_right (fun v g -> @v add_vertex g v) vs empty_graph)
\end{lstlisting}
  \caption{An excerpt of the interface of our graph library. A \code{('v, 'e) graph}
    represents a graph with vertices of type \code{'v} and edges
    (weights or labels) of type \code{'e}. Note how the
    \code{add\_vertex} and \code{add\_edge} operations in the
    implementation of \code{new\_graph} are scaled by cost variables
    \code{@v} and \code{@e} respectively.}
  \label{fig:graph-types}
\end{figure}

We have implemented two different functional graphs based on maps:
the first emulates an adjacency matrix by mapping pairs of vertices
$(a, b)$ to the edge (if any) connecting $a$ to $b$, while the second
maps each vertex $a$ to its adjacency list---a list of edge-node pairs representing $a$'s outgoing edges.
In the first implementation, the cost of getting all the outgoing edges of a
vertex is proportional to the number of vertices in the graph. 
In the second implementation, the
same operation is logarithmically proportional to the number of vertices in the graph (finding and returning the adjacency list).
On the other hand, querying the presence of an edge between two vertices is cheaper with an adjacency matrix since the full adjacency list needs to be traversed in order to find all outgoing edges.



While it would be possible to implement graphs using the universal
collection types from Section~\ref{sec:evaluation:case-uct}, we have
not done so here for technical reasons. Part of the issue is that we
currently do not support relating variables used for different
libraries, e.g., the collection length \code{n} and the number of
vertices and edges \code{v} and \code{e}.
For simplicity, we also do not differentiate graph implementations
based on properties, such as the order and retention properties of
universal collection types. One could imagine graph properties such as
being (un)directed or having weights.
These are all directions for future work.

\subsection{Claim~\ref{claim:extensible}: Extensibility}\label{sec:claim:extensibility}

The core design of our approach immediately enables extensibility
without changing a previously defined library on three axes: adding
\reprs (via \code{letrepr}), \operations (via \code{letop}), and
\impls (via \code{letimpl}). However, there are still points worth
mentioning, in particular what the consequences are of orthogonal
extensions. For example, if one library adds a new \operation, call it
\code{foo}, and another adds a new \repr, call it \code{bar\_r}, can
\code{foo} be used with \code{bar\_r}?

The answer depends on the design of the library. In the libraries
presented in this paper we use a design where all new \operations must
have at least one default \impl, written in terms of a closed set of
fundamental \operations. In this design \code{bar\_r} would have an
\impl of \code{foo}, even though neither extending library has
knowledge of the other, as long as there are enough fundamental
\operations for \code{bar\_r}. However, even if it did not, the end
result would typically not be a major issue; \code{bar\_r} would still
be a valid \repr choice in programs, merely not when and where
\code{foo} is used.

When considering extensibility on the level of \abstractions there are
two possible interpretations: adding new entirely independent
\abstractions, and adding new variations of pre-existing
\abstractions. The former is clearly possible by adding a new type
alias over a \code{repr} type parameterized with some newly defined
type. On the other hand, adding a new variation of a pre-existing
\abstraction requires intentional design and support in the original
library. For example, the UCT library presented in
Section~\ref{sec:evaluation:case-uct} is parameterized by properties
that dictate the semantics of a collection. With such a design we
could, e.g., add a bidirectional map as follows:

\begin{lstlisting}[language=RepCaml]
type keep_last_key_and_value
type ('k, 'v) bi_map = ('k * 'v, keep_last_key_and_value * _) coll
\end{lstlisting}

\noindent This requires no modification of the original library, and
new \reprs and \impls can now use this new property to provide
concrete semantics.

\subsection{Claim~\ref{claim:repr-switching}: Representation Switching}\label{sec:claim:repr-switching}

Our approach supports representation switching as a consequence of
allowing the type signature of a \code{letimpl} to mention multiple
distinct \reprs. This also means that \repr switching can only happen
where there is an \opUse, and only if that \operation supports
it. Broadly speaking there are three ways this might manifest, which
we exemplify here using the UCT library.

First, a library may provide an \operation whose express purpose is to
convert between \reprs. The UCT library has this in the form of the
\code{view} \operation, which can switch \reprs as well as
properties. It is defined as follows:

\begin{lstlisting}[language=RepCaml]
letop view : ('a, 'p1) coll -> ('a, 'p2) coll
letimpl[0.0] view : ('a, 'p) coll -> ('a, 'p) coll = fun x -> x;;
letimpl[1.0] view = foldl (@n append) empty;;
\end{lstlisting}

\noindent Note that there are two default implementations, one that
preserves properties and \repr and costs nothing, and one that builds
a new collection through \code{foldl}, \code{append}, and
\code{empty}. Each \repr can of course also provide specialized \impls
for conversions that can be made more efficiently.

Second, an \impl may explicitly switch \reprs in an \operation whose
primary purpose is something else. For example, our library contains
an \impl of \code{map} that converts between a cons-list and a
snoc-list using \code{List.rev\_map}, which is a tail-recursive
\code{map} that produces a reversed list as output. The relevant
\code{letimpl}s can be seen below.

\begin{lstlisting}[language=RepCaml]
letimpl[n] map : _ -> !list_r -> !snoc_r = List.rev_map;;
letimpl[n] map : _ -> !snoc_r -> !list_r = List.rev_map;;
\end{lstlisting}

\noindent Finally, a default \impl may allow \repr switching through
particular choices of internal \impls. For example, the default \impls
for \code{map} use \code{empty}, \code{append} (resp. \code{prepend}),
and \code{foldl} (resp. \code{foldr}). Importantly, one \repr choice
can be made for \code{foldl} or \code{foldr}, and another for
\code{empty} and \code{append} or \code{prepend}.

\begin{lstlisting}[language=RepCaml]
letimpl[1.0] map = fun f xs ->
  foldl (fun acc x -> @n append acc (f x)) empty xs;;
letimpl[1.0] map = fun f xs ->
  foldr (fun x acc -> @n prepend (f x) acc) empty xs;;
\end{lstlisting}

\noindent For example, \code{(map Char.code)} (where \code{Char.code}
converts from a character to an integer representing its ASCII
codepoint) can be used to convert from \code{string\_r} to
\code{list\_r} by picking an \impl using \code{string\_r} for the fold
and \impls using \code{list\_r} for the other two \operations.

This means that \repr switching is often available even without
explicit support for it. However, there are still two notable
limitations to this approach for \repr switching. First, \reprs can
only be switched at an \opUse, which might require, e.g., a manually
inserted call to \code{view}. Second, only statically available
information can be used to determine when to switch. Of course, this
is a trade-off: we incur no runtime overhead, but can also not adapt
to exceptional runtime behaviour, e.g., if collection sizes are wildly
different from what the programmer provided at compile-time.

\subsection{Claim~\ref{claim:selection-flexibility}: Flexible Representation Selection}\label{sec:claim:selection-flexibility}

Our approach does not require that all \reprs for any given
\abstraction are fully semantically equivalent, and will happily pick
\reprs that have correct semantics only for the \operations actually
used, even though they may behave drastically different for other
\operations, as long as the library is designed for it. For example,
the UCT library in Section~\ref{sec:evaluation:case-uct} puts semantic
properties in the type of a collection, and allows a \code{letimpl} to
ignore a property if it does not impact the semantics of the chosen
operation.

As a concrete example, a program might use the \code{ordered\_set}
type, which has the properties \code{keep\_last} and
\code{order\_seq}, which means that duplicate elements are discarded,
and elements are stored in insertion order. Yet if the program
never observes the order, e.g., if the only operations used are
\code{empty}, \code{append}, and \code{mem} (which checks for
membership), then the compiler need not actually pick an ordered set.

Concretely, this is because the relevant \impl definitions use an
underscore in place of the order property, since it does not matter to
the semantics of the function. For example, \code{rbset}, which is
backed by a red-black tree and thus in sorted order, defines
\code{empty}, \code{append}, and \code{mem} as follows:

\begin{lstlisting}[language=RepCaml]
letimpl[0.0] empty : !rbset_r = (* ... *)
letimpl[log n] append : !rbset_r ((_, keep_last * _) coll) -> _ -> !rbset
  = (* ... *)
letimpl[log n] mem : _ -> !rbset_r ((_, keep_last * _) coll) -> _
  = (* ... *)
\end{lstlisting}

\noindent Of course, this is only possible because of the design of
the UCT library and careful annotation of \impls, i.e., our approach
enables this property, but does not guarantee it.

\sectionBreakMaybe

\section{Compiler and Quantitative Evaluation}\label{sec:evaluation}

To further evaluate our approach we implement a compiler for a
sizeable subset\footnote{The subset encompasses most constructs related to pure functional programming. This includes, e.g., functions, algebraic data types, and pattern matching, but not mutability, objects, or exceptions.} of the OCaml programming language, which we then
extend with \code{repr} types, \code{letrepr}, \code{letop}, and
\code{letimpl}. The compiler, which we call \code{mi-ml}, is implemented in the Miking~\cite{bromanVisionMikingInteractive2019} framework and works by translating OCaml code to MCore, the core language of Miking. All processing is done on the MCore representation, after which we emit equivalent OCaml code, to be compiled by \code{ocamlopt}, the canonical OCaml compiler. Our compiler consists
of $\sim$1700 lines of MCore, along with
$\sim$8000 lines of \code{repr} related code, which we have added to
the Miking standard library. We make two claims relevant to this
section:

\begin{claims}
\item\label{claim:solver-flexibility} Our implementation enables efficient compilation during iterative development, as well as more
  expensive compilation yielding more efficient compiled programs when
  desired.
\item\label{claim:faster} Programs using our approach are typically
  faster than programs using idiomatic representation choices,
  especially when a more precise solver is used.
\end{claims}

\noindent All benchmarks were run on a laptop running NixOS 24.05 with
an Intel Core i7-8565U (4 cores) and 16GB of RAM. Measurements were
collected using \code{hyperfine}~\cite{Peter_hyperfine_2023} (for whole
program execution) or recording wall-time in the program itself (for
measuring time taken by each solver during compilation). In
total, the benchmarks took $\sim$13\unit{h}.

We cover Claim~\ref{claim:solver-flexibility} by measuring analysis
time and \solution cost for each of our solvers on randomly generated
programs with varying numbers of \opUses. Random program generation is
described in Section~\ref{sec:evaluation:random-programs}, analysis
time in Section~\ref{sec:evaluation:compile-overhead}, and solution
costs in Section~\ref{sec:evaluation:heuristic-precision}.

We cover Claim~\ref{claim:faster} by benchmarking a number of
handwritten programs, each with two variants: one using idiomatic
\repr choices, and one using \code{repr} types. The programs,
experimental setup, and analysis can be found in
Section~\ref{sec:evaluation:performance}.

\subsection{Randomly Generating Programs}\label{sec:evaluation:random-programs}

Each generated program follows a fairly simple pattern. There is a
code skeleton creating collections (using the UCT library of
Section~\ref{sec:evaluation:case-uct}) of various given sizes (read
from the program arguments), followed by a benchmark runner, setup to
run a particular function a given number of times (again read from the
program arguments) and print the elapsed wall-time. The skeleton has
two points where we add code: the initial type of the collection being
created, and the body of the function under test. Both of these are
filled in before compiling each program.

The initial collection type can be \code{seq} (i.e., a \code{coll}
with properties \code{keep\_all} and \code{order\_seq}), \code{set}
(properties \code{keep\_last} and \code{\_}), or \code{map}
(properties \code{keep\_last\_key} and \code{\_}).

The function body is a randomly selected sequence of code blocks, each
of which contains exactly one \operation. For example, below are the
code blocks containing \code{concat}, \code{split\_first}, and
\code{size}, respectively:

\begin{lstlisting}[language=RepCaml]
let coll = concat coll coll in
\end{lstlisting}

\begin{lstlisting}[language=RepCaml]
let coll =
  begin match split_first coll with
  | Some (_, coll) -> coll
  | None -> coll
  end in
\end{lstlisting}

\begin{lstlisting}[language=RepCaml]
Sys.opaque_identity (size coll);
\end{lstlisting}

\noindent Notably, each of these can appear in any order, and may or
may not shadow the collection (\code{coll}) with a new
value. \Operations that do not produce a new collection
(e.g. \code{size}) are wrapped in \code{Sys.opaque\_identity}, which
prevents them from being optimized away even if they have no
side-effects and their results are unused.

Note also that the set of possible \operations includes the
\code{view} \operation with an explicitly annotated return type,
whereby a generated program may switch collection semantics part-way
through the program. For example, a generated program starting out
with \code{seq} switches to \code{map} if it includes this \operation:

\begin{lstlisting}[language=RepCaml]
let coll : (int, int) map = view coll in
\end{lstlisting}

\noindent Finally, note that the initial collection is created outside
the function to be benchmarked, to capture interactions between other
\operations instead of always including the initial
creation. Similarly, the \operation used to create the collection is
given an explicit scaling factor of \code{0.0} to ensure that the
benchmarking and our analysis are concerned with the same part of the
program.

\subsection{Compile-time Overhead}\label{sec:evaluation:compile-overhead}

\newcommand{\initialCollTypes}{\code{seq}, \code{set}, and \code{map}}
\newcommand{\numProgPer}{5}
\newcommand{\maxOpCount}{40}
\newcommand{\nValues}{10, 100, 200, 1000, 2000, 5000, and 10000}

\newcommand{\exhaustiveAboveMinute}{686}
\newcommand{\transferAboveMinute}{0}
\newcommand{\bottomupAboveMinute}{217}
\newcommand{\greedyAboveMinute}{157}
\newcommand{\mixedAboveMinute}{0}
\newcommand{\zThreeAboveMinute}{782}
\newcommand{\slowestHeuristic}{1105.147949}

\begin{figure}
  \input{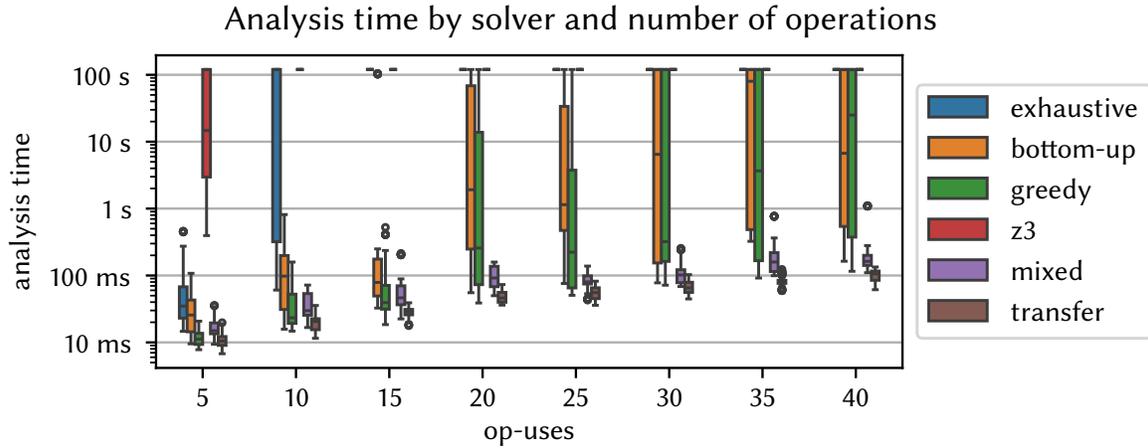}
  \caption{Box plots of the time taken to analyze programs and find
    \reprs and \impls, plotted by number of \opUses in each program
    and solver. Each box shows the median as a middle line, quartiles
    by top and bottom of the box, $1.5$ times the interquartile range
    by the whiskers, and outliers (data points appearing outside the
    whiskers) as circles. Timeouts are reported as though they took
    the full allowable time, i.e., 2 minutes. Note that \code{z3} times out for all inputs with more than 5 \opUses, and that the y-axis uses log-scale.\label{fig:evaluation:analysis-time}}
\end{figure}

The evaluation of our solvers, both total and heuristics-based, is
split over two sections: this section covers the amount of time needed
in relation to the number of \opUses, while the next examines the
cost of the produced \solutions (c.f. Fig.~\ref{fig:solver-types-and-cost} in
Section~\ref{sec:optimization-problem} for our definition of cost).

The data in both sections is obtained using the randomly generated
programs of Section~\ref{sec:evaluation:random-programs}, by
generating \numProgPer{} programs per size and initial collection type
(one of \initialCollTypes). We compile each program multiple times, each with a different value of \code{n} (\nValues). We limit the
execution time of each run of the compiler to 2 minutes.

Fig.~\ref{fig:evaluation:analysis-time} shows the analysis time
required for each program and solver, plotted by number of \opUses in
the program. As a reminder from Section~\ref{sec:solving-the-problem}:
\code{exhaustive} finds all possible solutions, \code{bottom-up},
\code{greedy}, and \code{z3} are total solvers that find an optimal
solution if one exists, while \code{mixed} is heuristics-based, and
\code{transfer} extends \code{mixed} by being able to apply previous
\solutions to new programs. Note that the graph includes both
successful and failed runs, and that all failures were due to timeout. Timeouts are recorded as though they took precisely the
maximum allowable time, i.e., 2 minutes. Note also that the variance
grows as programs get bigger, especially for the total solvers; this
is largely due to some \operations having more possible solutions than
others, e.g., \code{empty} typically only has direct \impls, while
\code{map} has both direct \impls and multiple default
\impls. A program with, e.g., one \code{empty} and one \code{map} will
thus have a larger solution space than one with only two uses of
\code{empty}.

As Fig.~\ref{fig:evaluation:analysis-time} suggests,
\code{exhaustive} listing of \solutions quickly becomes infeasible;
$\geq$50\% of programs produce a timeout already at 10 \opUses. The
other total solvers avoid the timeout for longer, but still have a
number of programs for which solving is slow, e.g., \code{bottom-up}
takes more than one minute for \bottomupAboveMinute{} programs. In
contrast, the heuristics-based solvers take at most
\num[round-mode=places,round-precision=0]{\slowestHeuristic}\unit{ms}.

Of course, the time a heuristic needs is not all that determines its
usefulness; it must also produce good results, which in our case means
programs whose cost is reasonably close to the optimal cost.

\subsection{Heuristics and Low-cost Programs}\label{sec:evaluation:heuristic-precision}

\newcommand{\worstFactorBottomUp}{1.0}
\newcommand{\worstFactorExhaustive}{1.0}
\newcommand{\worstFactorGreedy}{1.0}
\newcommand{\worstFactorMixed}{7.090407138909113}
\newcommand{\worstFactorTransfer}{7.090407138909113}
\newcommand{\worstFactorZ}{1.0}

\begin{figure}
  \input{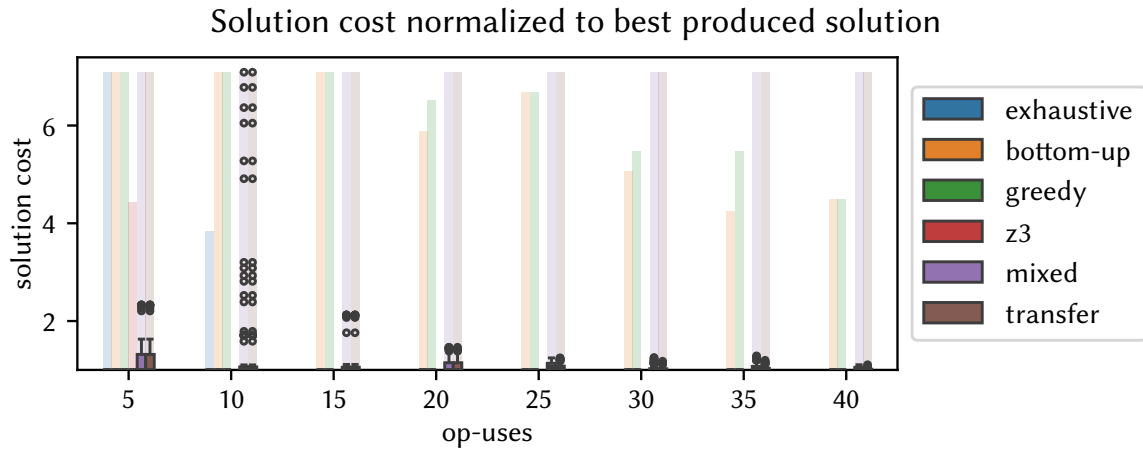}
  \caption{The cost of \solutions produced by each solver, normalized
    to the lowest cost \solution produced by any solver. Note that
    while \code{exhaustive} produces multiple solutions, the plot only
    uses the one with the lowest cost. Note also that timeouts are
    excluded; the faded bars in the background show the fraction of
    solutions that completed in time. This means that the total solvers
    are always recorded as optimal, even though they fail to produce
    \solutions for many larger inputs.\label{fig:factor-optimal}}
\end{figure}

In this section we examine the quality of \solutions produced by
our solvers. For each program, value of \code{n}, and solver, we record the cost (as defined in Fig.~\ref{fig:solver-types-and-cost}) of the produced \solution, ignoring runs that did not finish in time. Next, we normalize each cost to the lowest cost produced for the corresponding program and \code{n} by \emph{any} solver; in other words, we compare each solver against the best known \solution. The results can be seen in Fig.~\ref{fig:factor-optimal}. Note that for programs where at least one total solver (\code{exhaustive}, \code{bottom-up}, \code{greedy}, or \code{z3}) finished in time,\footnote{The total solvers begin using prohibitively much memory for larger inputs, thus simply running them for longer may not be enough to produce an optimal \solution for comparison.} the best known \solution is also the actual best \solution. The faded bars in the background show the fraction of runs that finished in time for each solver, and thus also what fraction of runs are represented in the figure.

Note that, as expected, the total solvers are always optimal. Note also that the highest normalized cost produced is $\sim$\num[round-mode=places,round-precision=2]{\worstFactorMixed},
which is lower than all values of \code{n} tested. Since the cost
annotations used in our UCT library are strongly connected to
asymptotic time complexity, this implies that the heuristic solvers (\code{mixed} and \code{transfer}) to a large
extent get the asymptotics correct, even if the solution produced is
not truly optimal. We observe that \code{transfer} produces better
\solutions than \code{mixed} for some larger inputs. This is because
it does not precisely re-apply a previous \solution; instead, it keeps
potential sub-\solutions as long as they appeared \emph{somewhere} in the
previous \solution, a sort of ``cross-pollination'', which sometimes
enables cheaper \solutions.

Finally, note that the set of possible \solution
costs is non-continuous. We could, e.g., construct a program where the
optimal solution has cost \code{1.0} while the second-best solution
has cost \code{1000.0}. This means that the magnitudes of the factors
in Fig.~\ref{fig:factor-optimal} are very much dependent on the
\impls available and the design of the library in which they are
defined.

Next, we examine the effect of our approach on the performance of
compiled programs.

\subsection{Performance Comparison}\label{sec:evaluation:performance}

When considering the performance of programs using our approach to
normal OCaml programs we use a number of handwritten programs,
with the intent that these programs are more representative of
programs in general than the randomly generated programs of the
previous subsections.

Each input program comes in two flavors: one using idiomatic \repr
choices, and one using \code{repr} types. We compile these to four executables: \code{ocamlopt} and \code{mi-ml} are the idiomatic version compiled with \code{ocamlopt} and \code{mi-ml}, while \code{mixed} and \code{transfer} are the \code{repr} version compiled with \code{mi-ml} and the corresponding solver. The \code{repr} versions are parameterized
with $\code{n}=\code{v}=\code{e}=\code{10000.0}$.

\begin{table}
  \sisetup{round-mode=places,round-precision=2}
  \begin{tabular}{l@{\qquad}S[table-format=1.2]S[table-format=1.2]S[table-format=1.2]S[table-format=1.2]@{\qquad} S[table-format=1.2] S[table-format=1.2]S[table-format=1.2]}
    & \multicolumn{4}{@{}c@{}}{Compilation time (s)} & \multicolumn{3}{@{}c@{}}{Running time (s)} \\
    Program & {\code{ocamlopt}} & {\code{mi-ml}} & {\code{mixed}} & {\code{transfer}} & {\code{ocamlopt}} & {\code{mi-ml}} & {\code{mixed}} \\ \midrule
    \code{show\_seq} & 0.1031281557586207 & 0.3698409609 & 0.8205291407 & 0.8052032303000001 & 10.0646882492 & 9.4010741532 & 0.03169478376923077 \\
\code{prepend} & 0.11913752832 & 0.4353684372 & 1.0333139308 & 0.9436329719 & 0.021577629903703703 & 0.021823853466666668 & 0.022008563574468083 \\
\code{append} & 0.12001915016666669 & 0.4337816832 & 1.0528795061 & 0.9337825243000003 & 186.43660080299998 & 178.4087547037 & 0.020285862596153847 \\
\code{pattern} & 0.2279931343846154 & 0.6697555582 & 3.6524520521999997 & 2.6081306905 & 10.0973200763 & 9.638234469 & 0.15276939757894736 \\
\code{dijkstra} & 0.15248739378947368 & 0.4695094813000001 & 0.8060618022 & 0.8060298895000001 & 6.442841586 & 7.7882150946 & 7.9163011998 \\
\code{celebrity} & 0.21353814241176472 & 0.48390961590000003 & 0.9395005531999999 & 1.0369345912 & 19.293998912000003 & 21.3556607924 & 11.967881688899999 \\

  \end{tabular}
  \caption{Time spent compiling and running handwritten test cases in
    four versions: code with no \code{repr} types compiled with \code{ocamlopt} and \code{mi-ml}, and code using
    \code{repr} types compiled with the \code{mixed} and
    \code{transfer} solvers. All measurements are end-to-end and
    include the entire execution time of the compiler or compiled
    program. We omit running time for \code{transfer} as the produced
    \solutions are identical to those of
    \code{mixed}.\label{tab:manual-performance}}
\end{table}

Table~\ref{tab:manual-performance} shows the results, both in terms of
compilation time and running time. Note that for these programs
\code{mixed} and \code{transfer} produced identical \solutions, thus
we omit the running time for \code{transfer}. Remember that our
compiler works by emitting OCaml code, thus the compilation time for
\code{mi-ml}, \code{mixed}, and \code{transfer} include running
\code{ocamlopt} in the background, i.e., we do not expect to have
truly competitive compilation times. Generally speaking, \code{mixed}
and \code{transfer} compile slower than \code{mi-ml}, since they do
more work, but produce executables that are as fast or faster,
sometimes by orders of magnitude. Next, we consider each program in
turn.

\paragraph{\code{show\_seq}}
This is the code from Fig.~\ref{fig:seq-ex:user} in
Section~\ref{sec:approach:user}, except using our UCT library, along
with surrounding code to build a sequence and print the resulting
string. The sequence is initialized using \code{List.init} or the
corresponding UCT \operation, depending on the program version. The
\code{repr} version is roughly three orders of magnitude faster here,
because the OCaml \code{string} type is backed by a byte array for
which concatenation requires copying both inputs (i.e., linear time),
while the \code{repr} versions select a
rope~\cite{boehmRopesAlternativeStrings1995} specialized to sequences
of characters, for which concatenation can be done in constant time.

\paragraph{\code{prepend} \& \code{append}}
These programs build a sequence by repeatedly \code{prepend}ing or
\code{append}ing an element, then print the result. Note that the
printing code uses \code{String.cat} (or the corresponding UCT
\operation), which efficiently concatenates a list of strings with a
delimiter between each element, rather than repeated concatenation of
two strings, as in \code{show\_seq}. The \code{list} type of OCaml
is a cons-list, i.e., \code{prepend} is efficient while \code{append}
is \emph{very} inefficient. The \code{repr} variants circumvent this
by using a snoc-list while building the sequence, i.e., a cons-list
stored in reverse, and then converting back to a normal
cons-list. This sort of juggling is common in language centered around
cons-lists, but can here be handled in the background by our
compiler. These two programs also illustrate how a small change in
\operation usage can drastically change the efficiency of a program
when the \repr is fixed, while our compiler can adapt without
additional source code changes.

\paragraph{\code{pattern}}
This program is a port of the pattern matching analysis machinery
present in the Miking compiler. The analysis handles regular patterns
(as in regular languages, i.e., ``and'', ``or'', and ``not''
patterns), as well as determining if two patterns overlap (i.e., there
are values that can be matched by both) or one is a subset of the
other (i.e., all values matched by one are also matched by the
other). The program computes and prints the complement of a large
randomly generated pattern (using a fixed seed, for
reproducibility). This program contains 383 \opUses, yielding an
initial \solution space of $\sim$$10^{389}$ potential
\solutions. Despite this, our heuristics finish within $\sim$3 and 2
seconds respectively, and produce a significantly more efficient
executable.

\paragraph{\code{dijkstra}}
This program implements Dijkstra's algorithm using the
graph library of Section~\ref{sec:evaluation:case-ugt}. Input is a
randomly generated graph (with a fixed seed, as for \code{pattern}), and the program finds the shortest paths from all nodes to all nodes. Since the OCaml
standard library contains no graph type there is no \emph{obvious}
idiomatic representation, instead we use the adjacency list \repr
described in Section~\ref{sec:evaluation:case-ugt}.
Dijkstra's algorithm gets the outgoing edges from each visited node, so the performance of this operation has a big impact on the overall performance. When using adjacency lists, these lists can be returned immediately, whereas with an adjacency matrix the outgoing edges need to be calculated (this makes the program take $\sim$14 seconds instead of the $\sim$7 seconds shown in Table~\ref{tab:manual-performance}). The \code{repr} variant selects the implementation based on adjacency lists.

\paragraph{\code{celebrity}}
This program implements the "Celebrity problem" that queries a graph for a node that has no outgoing edges but that every other node has an edge to. The program generates a random graph and repeats the search for celebrities 10,000 times (this is to reduce the relative time taken to create the graph). The algorithm repeatedly queries the graph for the existence of specific edges, something that is faster using an adjacency matrix than using adjacency lists. The \code{repr} variant selects the implementation based on adjacency matrices, meaning that it outperforms the implementation based on adjacency lists used by the idiomatic version.
\medskip

In conclusion, \code{mixed} finishes quickly for small programs and
within a reasonable amount of time for larger programs, producing
efficient executables for all of them.

\sectionBreakMaybe

\section{Related Work}\label{sec:related-work}

\newcommand{\appStatic}{Static}
\newcommand{\appBenchmark}{Bench}
\newcommand{\appDynamic}{Dynamic}
\newcommand{\appAll}{S+B+D}

\newcommand{\citeText}[1]{#1~}

\begin{table}
\begin{tabular}{rccccccc}
 & Approach & \multicolumn{4}{@{}c@{}}{Extensibility} & Switching & Flexibility \\
 & & Ty & Repr & Op & Impl & & \\ \midrule

\code{repr} (this paper)
& \appStatic
& \checkmark & \checkmark & \checkmark & \checkmark
& \checkmark
& \checkmark
\\ \addlinespace

\cite{lowAutomaticCodingChoice1976,lowAutomaticDataStructure1978}
& (\appBenchmark+)\appStatic
& & & &
&
& \checkmark
\\

\cite{anderssonProfileGuidedComposition2008}
& \appBenchmark+\appDynamic
& \checkmark & \checkmark & & \checkmark
& \checkmark
&
\\

\citeText{Chameleon}\cite{shachamChameleonAdaptiveSelection2009}
& \appBenchmark
& & \checkmark & &
&
&
\\

\citeText{PetaBricks}\cite{anselPetaBricksLanguageCompiler2009}
& \appBenchmark+\appStatic
& & & \checkmark &
& \checkmark
&
\\

\citeText{Brainy}\cite{jungBrainyEffectiveSelection2011,coudercClassificationbasedStaticCollection2023}
& \appBenchmark+\appStatic$^*$
& & & &
&
& \checkmark
\\

\citeText{CoCo}\cite{xuCoCoSoundAdaptive2013}
& \appDynamic
& \checkmark & \checkmark & &
& \checkmark
&
\\

\cite{osterlundDynamicallyTransformingData2013}
& \appDynamic
& \checkmark & & &
& \checkmark
&
\\

\cite{kusumAdaptingGraphApplication2014,kusumSafeFlexibleAdaptation2016}
& \appDynamic
& \checkmark & \checkmark & &
& \checkmark
&
\\

\citeText{Late Data Layout}\cite{urecheAutomatingAdHoc2015}
& \appStatic$^\dagger$
& \checkmark & \checkmark & &
& \checkmark
&
\\

\citeText{SimpleL}\cite{binCodeGenerationAbstract2015}
& \appStatic
& & & &
&
& \checkmark
\\

\citeText{JitDS}\cite{dewaelJustintimeDataStructures2015}
& \appDynamic
& \checkmark & & &
& \checkmark
& \checkmark
\\

\cite{schillerCompileRuntimeApproaches2016}
& \appAll
& & & &
& \checkmark
&
\\

\citeText{Artemis}\cite{basiosOptimisingDarwinianData2017,basiosDarwinianDataStructure2018}
& \appStatic+\appBenchmark
& \checkmark & \checkmark & &
&
&
\\


\citeText{DBFlex}\cite{shaikhhaFineTuningDataStructures2023}
& \appStatic+\appBenchmark
& & \checkmark & &
&
&
\\

\citeText{Cres}\cite{wangComplexityguidedContainerReplacement2022}
& \appStatic
& \checkmark & \checkmark & &
&
& \checkmark
\\

\citeText{CT+}\cite{oliveiraRecommendingEnergyEfficientJava2019,oliveiraImprovingEnergyefficiencyRecommending2021}
& \appBenchmark+\appStatic$^*$
& & \checkmark & &
&
&
\\

\citeText{SETL}\cite{schonbergAutomaticTechniqueSelection1981}
& \appStatic
& & & &
&
&
\\

\cite{cohenAutomatingRelationalOperations1993}
& \appStatic$^\dagger$
& & \checkmark & &
&
& \checkmark
\\ \addlinespace

\end{tabular}\\
$^*$ Analysis only \qquad $^\dagger{}$ No analysis

\caption{Overview of related work, highlighting fulfilment of the
  non-quantitative challenges from Section~\ref{sec:introduction}. We
  mark a challenge if a work mentions it explicitly or strongly
  implies support for it. Works without analysis provide an easier way
  to manually switch \reprs, but no automatic selection mechanism,
  while works with analysis only provide suggestions for change, but
  no automatic means of applying the changes.\label{tab:related-work}}
\end{table}

There is a significant body of work concerned with automatically
picking or creating efficient \reprs of data structures. We summarize
most of these in Table~\ref{tab:related-work}. The first column lists
the overarching approach and when decisions are made: \emph{static}
analysis with choices at compile-time, \emph{dynamic} analysis using
run-time data and making choices during run-time, and \emph{bench} for
those using data from benchmarks or previous executions. The remaining
six columns state which of the challenges we list in
Section~\ref{sec:introduction} are tackled by each work. Extensibility
is split in four columns: ``ty'' if new \abstractions can be defined,
``repr'' and ``op'' if previously defined \abstractions can be extended
with new \reprs and \operations, respectively, and ``impl'' if new
\impls can be added to previously defined \reprs and
\operations. ``Switching'' refers to \repr switching, where the \repr
of a value can change over its lifetime, while ``flexibility'' is for
approaches where \repr selection can be done based on observed
semantics, rather than requiring all \reprs to be fully semantically
equivalent.

Note that we add a checkmark if the corresponding paper explicitly
mentions or strongly implies a particular capability. This means that
the table may contain false negatives, e.g., if an approach supports
extension along a particular axis, but the paper authors did not
consider this a central property of their contribution. Note also that
not all related work fits neatly in Table~\ref{tab:related-work}; we
return to these at the end of this section.

\paragraph{Dynamic approaches}
These works collect some amount of data during run-time and have
varying approaches for making the decision to switch \repr, trading
some amount of run-time overhead for more usable data. For example,
note that all dynamic approaches support \repr switching while most
static approaches do
not. \citet{xuCoCoSoundAdaptive2013,dewaelJustintimeDataStructures2015}
both allow the user to specify the data and switching criteria with
varying level of compiler-support, while
\citet{osterlundDynamicallyTransformingData2013} is a bit more
structured; they use a user-specified finite automaton, where edges
are labelled with \operations and some states are labelled with \reprs
to switch to. \citet{anderssonProfileGuidedComposition2008} uses a
combination of benchmarking and estimation functions to pre-compute
\repr choices by context, e.g., the size of a collection, or the type
of its elements, then uses that information during
execution. \citet{kusumAdaptingGraphApplication2014,kusumSafeFlexibleAdaptation2016}
uses manual annotations stating which \repr is optimal at different
intervals of user-defined statistics, intended to be determined
through benchmarking. Note that these annotations are per-application
rather than per-operation, in contrast to most other
approaches. Finally, most of these keep at most one \repr per value at
a time, while \citet{xuCoCoSoundAdaptive2013} conceptually maintains
all valid choices at once, with significant optimizations to avoid the
overhead of having several redundant \reprs in memory at a time.

\paragraph{Static approaches}
These works leave no run-time overhead in exchange for less precise
data, and sometimes take advantage of the ability to run more costly
analysis or benchmarks in an offline approach compared to an online
one. Our approach fits in this category. Most static works have a cost
model where each \impl has an associated cost, obtained through manual
annotations~\cite{lowAutomaticCodingChoice1976,lowAutomaticDataStructure1978,wangComplexityguidedContainerReplacement2022}
or benchmarking of individual
\impls~\cite{jungBrainyEffectiveSelection2011,coudercClassificationbasedStaticCollection2023,schillerCompileRuntimeApproaches2016,shaikhhaFineTuningDataStructures2023,oliveiraRecommendingEnergyEfficientJava2019,oliveiraImprovingEnergyefficiencyRecommending2021}. Several
of these allow using data from a previous instrumented run to direct
the analysis, typically by weighting the importance of each \operation
by either invocation count or execution
time~\cite{lowAutomaticCodingChoice1976,lowAutomaticDataStructure1978,jungBrainyEffectiveSelection2011,coudercClassificationbasedStaticCollection2023,oliveiraRecommendingEnergyEfficientJava2019,oliveiraImprovingEnergyefficiencyRecommending2021}. \citet{lowAutomaticCodingChoice1976,lowAutomaticDataStructure1978,wangComplexityguidedContainerReplacement2022}
have cost models quite similar to ours, though theirs are symbolic
(i.e., costs have a partial order and are not merely floating point
numbers), and, as far as we can tell, they do not handle the recursive
formulation where \impls can use other
\impls. \citet{wangComplexityguidedContainerReplacement2022} in
particular uses a more granular form of the properties we discuss in
our UCT library (c.f. Section~\ref{sec:evaluation:case-uct}) where
\impls may mutate or read properties, and \reprs are chosen based on
observable effect on these
properties. Brainy~\cite{jungBrainyEffectiveSelection2011,coudercClassificationbasedStaticCollection2023}
builds a machine learning model to select \reprs, while
Artemis~\cite{basiosOptimisingDarwinianData2017,basiosDarwinianDataStructure2018}
uses a genetic algorithm combined with
benchmarking. PetaBricks~\cite{anselPetaBricksLanguageCompiler2009}
focuses on \operations and \impls, permitting far greater granularity
in composition, e.g., automatically switching \impls depending on
input size, including in recursive calls from other \impls, or
decomposing the input and running separate \impls on each part. Note
also that PetaBricks and our approach are the only approaches in
Table~\ref{tab:related-work} that support \repr switching without a
dynamic component.

\paragraph{Other approaches}
These works address the difficulty of picking \reprs from a different
angle, opting to control state in a larger sense, making it easier to
manually make changes, or simply assisting in detecting suboptimal
implementation
choices. RelC~\cite{hawkinsDataRepresentationSynthesis2011} allows a
user to specify application state using relational algebra along with
a decomposition into conventional collections, as well as a set of
desired queries and state updates. The decomposition can be
automatically found and tuned, and code implementing queries and
updates is generated. Costs are obtained via
benchmarking. Cozy~\cite{loncaricFastSynthesisFast2016,loncaricGeneralizedDataStructure2018}
has a similar goal, but drops the explicit decomposition, allows
relating multiple collections, and uses an explicit cost model over
its internal language rather than
benchmarking. \cite{xuDetectingInefficientlyusedContainers2010,xuFindingLowutilityData2010}
both detect inefficiently used containers, where the cost of
creating the container outweighs its benefit, either by counting
writes vs. reads, or by tracking reachability in a data-dependency
graph. Clarity~\cite{olivoStaticDetectionAsymptotic2015} finds
collections that are traversed multiple times without intermediate
mutation, i.e., where some traversals could be merged and/or reuse
previously computed
data. Chameleon~\cite{shachamChameleonAdaptiveSelection2009} collects
data at run-time, then processes it offline to produce suggested \repr
changes based on manually written rules, no automatic changes are
made. Late Data Layout~\cite{urecheAutomatingAdHoc2015} allows a user
to specify type substitutions along with optimized methods, which can
then be used to pick a different \repr per program scope using manual
annotations.

\sectionBreakMaybe

\section{Conclusion and Future Work}

In this paper we introduce an approach that allows a programmer to
define \abstractions for which the compiler picks \reprs that is
extensible, supports \repr switching, and allows selecting \reprs
based on observable semantics. We incorporate our approach in a
compiler for a sizeable subset of the OCaml programming language,
which we then use to define a universal collection type as a library,
i.e., one type that represents all conventional collections through
different semantic properties, as well as a representation-flexible
graph library. Our compiler includes multiple solvers with different
trade-offs between compilation time and optimality of the compiled
program. We evaluate the solvers by measuring compilation time as well
as running time on a number of randomly generated as well as
handwritten programs, showing good results on both compilation and
running time.

There are several avenues of future work that seems promising. First,
while our implementation exploits the independence of sub-problems, it
does not exploit repeated patterns within or between sub-problems. In
practice, most programs tend to have a large degree of repeated
structure, thus we expect this to further improve scalability. Second,
we have designed our type system additions and implementation
selection to preserve well-typedness, and while this appears to hold
in practice, a formal proof is desirable. Finally, we would like to
expand our prototype such that we can conduct more experiments on
larger, pre-existing software projects.

This work was supported by the Vinnova Competence Center for
Trustworthy Edge Computing Systems and Applications (TECoSA) at the
KTH Royal Institute of Technology, Digital Futures, and the Swedish
Research Council (Vetenskapsrådet, Grants No. 2020-05346, 2023-05526,
and 2018-04329).

\sectionBreakMaybe

\bibliographystyle{ACM-Reference-Format}
\bibliography{All,Manual}

\sectionBreakMaybe

\end{document}


\appendix

\section{Checking the validity of a \solution}

\begin{algorithm}[t]
    Check that $\project{\mOpUse}{\mOp} = \projectTwo{\mSol}{\mImpl}{\mOp}$\Comment*{\boxed{\Validate{$\mSol$, $\mOpUse$}}}\label{line:validate:op-check}

    \SetVar{\mType}{\projectTwo{\mSol}{\mImpl}{\mType}}\;

    \SetVar{\many{\mOpUse}}{\projectTwo{\mSol}{\mImpl}{\many{\mOpUse}}}\;

    Instantiate $\mType$ and perform the corresponding substitutions
    throughout $\many{\mOpUse}$ as well. For example, a type \code{'a
      -> 'a} is instantiated by generating a fresh unification
    variable, call it $u$, and replacing \code{'a} with $u$, resulting
    in $u$\code{ -> }$u$\;\label{line:validate:instantiate}

    For each local \reprVar (i.e., a \reprVar ocurring only inside the
    \code{letimpl} corresponding to $\project{\mSol}{\mImpl}$),
    replace it with a fresh \reprVar throughout $\mType$ and
    $\many{\mOpUse}$\;\label{line:validate:fresh-local-reprs}

    Repeat~\ref{line:validate:fresh-local-reprs} for local (type)
    unification variables\;\label{line:validate:fresh-local-types}

    Record \repr assignments in $\mType$, e.g., for a type
    \code{!rep~(T~repr)} where \code{T~repr} has the \reprVar $r$,
    record $r = \code{rep}$\;\label{line:validate:repr-assignment}

    Remove \repr assignments in $\mType$, e.g., replace
    \code{!r~(T~repr)} with \code{T~repr}\;

    Substitute a fresh unification variable for each underscore
    present in $\project{\mOpUse}{\mType}$, then unify the resulting
    type with $\mType$\;\label{line:validate:unify}

    \lFor{$i$ \In $1$ \KwTo
      $\length{\many{\mOpUse}}$}{\Validate{$\getAt{\project{\mSol}{\many{\mSol}}}{i}$,
        $\getAt{\many{\mOpUse}}{i}$}}\label{line:validate:recursive-call}
  \caption{\protect\Validate{$\mSol$, $\mOpUse$} checks that $\mSol$
    is a valid \solution for a given \opUse $\mOpUse$. There are four
    possible reasons for a \solution to be invalid: i) the chosen
    \impl is for another \operation
    (step~\ref{line:validate:op-check}), ii) one \reprVar is given
    multiple \reprs, possibly across different recursive calls to
    \protect\Validate (step~\ref{line:validate:repr-assignment}), iii)
    the types are incompatible (line~\ref{line:validate:unify}), and
    iv) a sub-\solution is invalid
    (step~\ref{line:validate:recursive-call}). Steps~\ref{line:validate:fresh-local-reprs}
    and~\ref{line:validate:fresh-local-types} ensure that multiple
    instances of the same \impl in a \solution do not
    conflict.\label{alg:validate}}
\end{algorithm}

Intuitively speaking, a \solution is \emph{valid} if the corresponding
transformed program is type-safe and each \reprVar has at most one
\repr\footnote{A \reprVar that is irrelevant to the program need not
be assigned a \repr.}. Concretely a $\mSol$ is valid for a given
$\mOpUse$ if the \Validate predicate, given in
Algorithm~\ref{alg:validate}, terminates without finding an
inconsistency. There are four checks that may fail:

\begin{itemize}
\item The chosen \impl must implement the \operation the \opUse refers
  to (step~\ref{line:validate:op-check}).
\item Each \reprVar must be given at most one \repr
  (step~\ref{line:validate:repr-assignment}).
\item The annotated type of the \impl must be consistent with the
  inferred type of the \opUse (step~\ref{line:validate:unify}). Note
  that this is checked via unification, which might unify \reprVars,
  which in turn may lead to one \reprVar being assigned multiple
  \reprs.
\item Each sub-solution must be valid for the corresponding \opUse in
  the chosen \impl (step~\ref{line:validate:recursive-call}).
\end{itemize}

\noindent For example, $\mathit{sol} := [(\mathit{contains}_2,
  [(\mathit{fold}_1, [])])]$ is a valid \solution for the single
\opUse in the running example because a call to
\Validate{$\mathit{sol}$, $(\code{1.0}, \code{contains}, \code{int ->
    int repr$_7$ -> bool})$} proceeds as follows:

\begin{enumerate}
\item The \impl is for the correct \operation: $\code{contains} =
  \code{contains}$ holds.
\item \SetVar{\mType}{(\code{'d -> 'd repr$_6$ -> bool})}
\item \SetVar{\many{\mOpUse}}{[(\code{1.0}, \code{fold}, \code{(bool -> 'd -> bool) -> bool -> 'd repr$_6$ -> bool})]}
\item We instantiate $\mType$, generating a fresh unification variable
  (call it $u$) and substituting it for every ocurrence of
  \code{'d}:\\
  \begin{tabular}{r@{ }c@{ }l}
    $\mType$ & $=$ & $(\code{$u$ -> $u$ repr$_6$ -> bool})$ \\
    $\many{\mOpUse}$ & $=$ & $[(\code{1.0}, \code{fold}, \code{(bool -> $u$ -> bool) -> bool -> $u$ repr$_6$ -> bool})]$ \\
  \end{tabular}
\item The \impl contains only one \reprVar (6), and it does not appear
  anywhere else, thus we generate a new \reprVar (8) and substitute it
  for the old:\\
  \begin{tabular}{r@{ }c@{ }l}
    $\mType$ & $=$ & $(\code{$u$ -> $u$ repr$_8$ -> bool})$ \\
    $\many{\mOpUse}$ & $=$ & $[(\code{1.0}, \code{fold}, \code{(bool -> $u$ -> bool) -> bool -> $u$ repr$_8$ -> bool})]$ \\
  \end{tabular}

  For this particular example the outcome would not change if we
  skipped this step, since $\mathit{contains}_2$ only occurs once in
  the \solution. However, if a \solution uses the same \impl multiple
  times this step is important; without it all such instances would
  share \reprVars, which would heavily constrain the \solution.
\item This example contains no unresolved (type) unification
  variables, thus this step does nothing.
\item[7\&8]
  $\mType$ contains no \repr assignments, thus these steps also do
  nothing.\addtocounter{enumi}{2}
\item $\mType$ contains no underscores, thus this step unifies
  \code{$u$ -> $u$ repr$_8$ -> bool} with \code{int -> int repr$_7$ ->
    bool}, which implies that $u = \code{int}$ and \reprVars $7$ and
  $8$ are equal, yielding the following state:\\
  \begin{tabular}{r@{ }c@{ }l}
    $\mType$ & $=$ & $(\code{int -> int repr$_7$ -> bool})$ \\
    $\many{\mOpUse}$ & $=$ & $[(\code{1.0}, \code{fold}, \code{(bool -> int -> bool) -> bool -> int repr$_7$ -> bool})]$ \\
  \end{tabular}
\end{enumerate}

\noindent The recursive call to \Validate{$(\mathit{fold}_1, [])$,
  $\getAt{\many{\mOpUse}}{1}$}
(step~\ref{line:validate:recursive-call}) proceeds in a fashion
similar to the above until
step~\ref{line:validate:repr-assignment}. The state at that point can
be seen below (using $u'$ and $u''$ to denote the unification
variables introduced in step~\ref{line:validate:instantiate}, and 9 as
the new \reprVar introduced in
step~\ref{line:validate:fresh-local-reprs}):

\begin{center}
  \begin{tabular}{r@{ }c@{ }l}
    $\mType$ & $=$ & $(\code{($u'$ -> $u''$ -> $u'$) -> $u'$ -> !list\_r ($u''$ repr$_9$) -> $u'$})$ \\
    $\many{\mOpUse}$ & $=$ & $[]$ \\
  \end{tabular}
\end{center}

\begin{enumerate}
\setcounter{enumi}{6}
\item There is one \repr assignment, so we record that \reprVar 9 has
  \repr \code{list\_r}.
\item Removing the \repr assignment updates $\mType$:\\
  \begin{tabular}{r@{ }c@{ }l}
    $\mType$ & $=$ & $(\code{($u'$ -> $u''$ -> $u'$) -> $u'$ -> $u''$ repr$_9$ -> $u'$})$ \\
  \end{tabular}
\item Next, unifying $\mType$ with \code{(bool -> int -> bool) -> bool
  -> int repr$_7$ -> bool} implies $u' = \code{bool}$, $u'' =
  \code{int}$, and that \reprVars 7 and 9 are equal, which by
  extension means that \reprVar 7 also has the \code{list\_r} \repr.
\item $\many{\mOpUse}$ is empty, thus no recursive calls are needed.
\end{enumerate}

\noindent In the end \Validate terminates without any checks failing,
thus $\mathit{sol}$ is a valid \solution for the \opUse ocurring
outside a \code{letimpl} in the running example.

\section{Solvers in more detail}

This section describes our approach to solving the problem from a
high-level, along with slightly more detail on each of the components
of our solvers. All of our solvers center around an abstracted tree
representation of the solution space. Each sub-tree is rooted by one
of three kinds of nodes:

\begin{description}
\item[And,] which represents a single \code{letimpl}, with one child
  per \opUse. A valid \solution for the sub-tree requires a valid
  \solution for each child.
\item[Or,] which represents the disjunction of all \code{letimpl}s in
  scope for a particular \operation, i.e., it has one child per
  \code{letimpl}. A valid \solution for the sub-tree only requires a
  valid \solution for one child.
\item[Single,] which represents a complete \solution for that
  sub-tree. These occur as partial solutions during solving, but are
  not present in the initial tree.
\end{description}

\noindent Each node also contains some metadata:

\begin{itemize}
\item The cost the node would contribute to a final \solution.
\item The constraints that must hold if the node is to be in the final
  \solution, e.g., ``these unification variables must be equal'' or
  ``this \reprVar must be \code{list\_r}''.
\item An overapproximation of the constraints for the sub-tree, namely
  the set of possible \reprs for each \reprVar assigned anywhere in
  the sub-tree.
\item Two sets of variables (both unification and \reprVars), one
  containing variables relevant inside the sub-tree, the other
  containing variables relevant both inside and outside the
  sub-tree. These are used for pruning; \solutions that differ only in
  variables irrelevant outside the sub-tree are equivalent, i.e., we
  can retain only one such \solution without affecting the solvability
  of a problem.
\end{itemize}

\noindent Finally, each node also contains all data required to
construct a program from a \solution. This data is abstracted, in the
sense that the solvers never interact with it directly, they only
manipulate the tree according to its logical structure.

For example, the abstracted tree for the running example from
Figure~\ref{fig:implementation:running-example} can be seen below:

\begin{center}
  \includegraphics{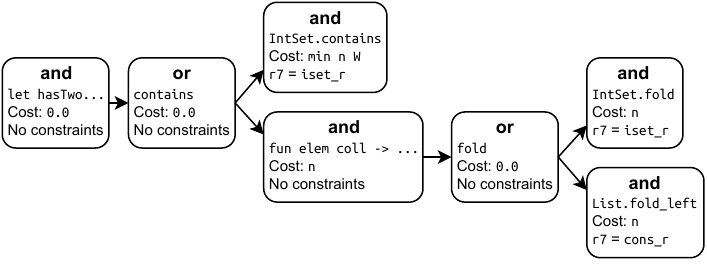}
\end{center}

\noindent Note that the root node of the tree is an \code{and} node
representing the entire program; it has one child per \opUse occurring
outside a \code{letimpl}, which in this case is just one:
\code{contains} in the body of \code{hasTwo}.

Constructing the tree (which happens before solving, regardless of the
chosen solver) is largely analogous to \Validate, with three important
changes:

\begin{itemize}
\item Instead of checking that an \impl is valid, we go through
  \emph{all} \impls for the given \operation to see which ones
  \emph{might} be valid, adding an \code{and} node for each of them,
  and making them children to a new \code{or} node. This means that
  our input is now all \impls in scope rather than one potential
  \solution, and the return is a tree, rather than the result of a
  number of checks.
\item The \repr assignment in step~\ref{line:validate:repr-assignment}
  and the unification in step~\ref{line:validate:unify} are no longer
  side-effecting, instead we store the resulting constraints in the
  corresponding \code{and} node. These constraints are then used to
  update the \opUses (i.e., they propagate down) but do not propagate
  up at this point. This is because we do not yet know that this \impl
  will be in the final \solution, thus its constraints may be
  irrelevant.
\item If we recurse to the same \operation with an \opUse whose
  inferred type is also the same (modulo renaming of variables) we
  return an empty \code{or} node, to avoid infinite recursion. This is
  a safe omission, because any \solution that recurses to the same
  \operation and type twice implies the existance of a \solution
  without the double recursion. Furthermore, given non-negative costs
  on all \code{letimpl}s, the latter \solution has an equal or lower
  cost, i.e., the \solutions we discard are never useful.
\end{itemize}

\noindent Importantly, all other steps are unchanged, e.g., we still
instantiate the type of each \impl and generate new unification and
\reprVars for variables that only occur locally.

We also implement a number of operations on the abstract tree, some of
which are shared across multiple solvers:

\begin{description}
\item[Flatten] finds \code{And} nodes that are direct children of other \code{And} nodes, then flattens them into a single \code{And} node.
\item[Propagate] propagates constraints across the tree and performs a
  number of simplifications:
  \begin{itemize}
  \item \code{And} nodes where all children are \code{Single} are
    transformed into new \code{Single} nodes.
  \item \code{Or} nodes with only one child are replaced by the child.
  \item Nested \code{or} nodes are flattened; the children of each
    nested \code{or} become children of the parent \code{or} instead.
  \item \code{Single} children of \code{or} nodes are pruned if they
    have neither a better cost nor a more flexible set of constraints
    than one of its \code{single} siblings.
  \item Nodes with contradictory constraints or where a \reprVar has
    an empty set of possible \reprs are removed.
  \end{itemize}
\item[Partition] looks at each \code{And}-node, noting which children interact with which unification and \reprVars, and then partitions them such that all partitions are independent of each other, there is no overlap in variables.
\item[Collapse leaves] finds \code{And} nodes where all children are
  \code{single} (or \code{Or} nodes with only \code{single} children)
  and replaces them with an \code{Or} node of all valid combinations.
\item[Homogenize] takes a sub-tree, notes the set of possible \reprs
  for each \reprVar, and then produces one alternative sub-tree for
  each maximally homogeneous assignment of \reprs. We consider an
  assignment to be maximally homogeneous if we can construct it by
  picking a \repr, assigning it to all \reprVars for which it is
  valid, then picking another \repr and assigning it where it is
  valid, etc., until all \reprVars have an assigned \repr.
\item[Materialize lazy] takes a sub-tree and produces a lazy sequence
  of all its valid \solutions, in cheapest-first order. Guaranteeing
  cheapest-first and never returning duplicates induces some overhead
  in tracking attempted combinations of children and potentially
  computing more \solutions for each child than we end up using, to
  guarantee that the new \solution cannot produce a cheaper
  combination.
\item[Materialize guided] attempts to produce \solutions for a given
  sub-tree by recursively obtaining a \solution for each of the
  children of an \code{And} node and checking if they are all
  consistent with each other. If they are this new \solution is
  returned, otherwise we compute a combined constraint from the
  consistent children, and recursively request new \solutions for the
  other children, limited to \solutions that are consistent with the
  combined constraint. This is repeated until a \solution is found, or
  the procedure fails.
\end{description}

\noindent All solvers will run \code{propagate}, \code{flatten}, and \code{partition} in that order before doing anything else.

We provide a total of 8 solvers; 1 solver that produces
\emph{all} valid \solutions, 3 total solvers that find an optimal
\solution if one exists, and 4 heuristics-based solvers that may fail
or produce a sub-optimal \solution for some inputs:

\begin{description}
\item[\code{exhaustive}] repeatedly applies \textbf{propagate}
  (without pruning redundant \solutions) and \textbf{collapse leaves}
  until the tree consists of an \code{or} node with \code{single}
  children.
\item[\code{bottom-up}] is the same as \code{exhaustive}, except
  redundant \solutions are pruned at every \code{Or} node. Note that
  no unification or \reprVars are relevant outside the tree, thus
  pruning of an \code{Or} node at the root finds all \solutions
  redundant except the cheapest one, if it exists, thus
  \code{bottom-up} returns at most one \solution.
\item[\code{greedy}] uses \textbf{materialize lazy} on the root of the
  tree and picks the first \solution, if one exists.
\item[\code{z3}] encodes the tree as an SMTLIB2 model which is then
  passed to Z3~\cite{demouraZ3EfficientSMT2008a} to be solved
  externally. The result is then parsed and applied to the tree to
  produce a final \solution.
\item[\code{homogeneous}] uses a variant of \textbf{homogenize} that
  only produces one maximally homogeneous variant, and then applies it
  recursively at every point in the tree, going top to bottom.
\item[\code{guided}] uses \textbf{materialize guided} on the root of
  the tree and then returns the produced \solution, if any.
\item[\code{mixed}] uses both \code{homogeneous} and \code{guided} to
  produce multiple \solutions and then picks the cheapest. We apply
  both solvers to the root of the tree, but also to each alternative
  produced by \textbf{homogenize} when applied to the root.
\item[\code{transfer}] filter an input tree to only consider options that were useful in some previous \solution, then runs \code{mixed} on the filtered input.
\end{description}

Note that two of these, \code{guided} and \code{homogeneous}, are not used directly in the paper, only as part of \code{mixed} (and by extension, \code{transfer}).

\bibliographystyle{ACM-Reference-Format}
\bibliography{All,Manual}